%%%%%%%%%%%%%%%%%%     PLAIN TEX FILE
%%%%%%%%%%%%%%%%%%
 %%%%%%%%%%%%%%%%%%  %%%%%%%%%%%%%%%%%%  %%%%%%%%%%%%%%%%%%  %%%%%%%%%%%%%%%%%%
 %%%%%%%%%%%%%%%%%%  %%%%%%%%%%%%%%%%%%  %%%%%%%%%%%%%%%%%%  %%%%%%%%%%%%%%%%%%
 %%%%%%%%%%%%%%%%%%  %%%%%%%%%%%%%%%%%%  %%%%%%%%%%%%%%%%%%  %%%%%%%%%%%%%%%%%%

 %%%%%%%%%%%%%%%%%%  hex macros for preprints, cm version %%%%%%%%%%%%%%
%                     (P. Ginsparg, last updated 9/91)
%                if confused, type `b' in response to query
%
%---------------------------------------------------------------------%
%% site dependent options:
%% \unredoffs and \redoffs define horizontal and vertical offsets
%% respectively for unreduced and reduced modes. \speclscape defines
%% the \special{} call that sets printer to landscape (sideways) mode.
%% from standard set below, leave uncommented as appropriate or redefine
%
%%% next 400dpi
%\def\unredoffs{} \def\redoffs{\voffset=-.31truein\hoffset=-.48truein}
%\def\speclscape{\special{landscape}}
%
%%% apple lw
\def\unredoffs{} \def\redoffs{\voffset=-.31truein\hoffset=-.59truein}
\def\speclscape{\special{ps: landscape}}
%
%%% qms lasergrafix:
%\def\unredoffs{} \def\redoffs{\voffset=-.4truein\hoffset=.125truein}
%\def\speclscape{\special{qms: landscape}}
%
%%% saclay A4 paper:
%\def\unredoffs{\hoffset-.14truein\voffset-.2truein}
%\def\redoffs{\voffset=-.55truein\hoffset=-.1truein} \def\speclscape{}
%
%---------------------------------------------------------------------%
%
\newbox\leftpage \newdimen\fullhsize \newdimen\hstitle \newdimen\hsbody
\tolerance=1000\hfuzz=2pt
\catcode`\@=11 % This allows us to modify PLAIN macros.
\def\bigans{b }
%\message{ big or little (b/l)? }\read-1 to\answ
\def\answ{b }
\ifx\answ\bigans\message{(This will come out unreduced.}
\magnification=1200\unredoffs\baselineskip=16pt plus 2pt minus 1pt
\hsbody=\hsize \hstitle=\hsize %take default values for unreduced format
\else\message{(This will be reduced.} \let\l@r=L
\magnification=1000\baselineskip=16pt plus 2pt minus 1pt \vsize=7truein
\redoffs \hstitle=8truein\hsbody=4.75truein\fullhsize=10truein\hsize=\hsbody
\output={\ifnum\pageno=0 %%% This is the HUTP version
  \shipout\vbox{\speclscape{\hsize\fullhsize\makeheadline}
    \hbox to \fullhsize{\hfill\pagebody\hfill}}\advancepageno
  \else
  \almostshipout{\leftline{\vbox{\pagebody\makefootline}}}\advancepageno
  \fi}
\def\almostshipout#1{\if L\l@r \count1=1 \message{[\the\count0.\the\count1]}
      \global\setbox\leftpage=#1 \global\let\l@r=R
 \else \count1=2
  \shipout\vbox{\speclscape{\hsize\fullhsize\makeheadline}
      \hbox to\fullhsize{\box\leftpage\hfil#1}}  \global\let\l@r=L\fi}
\fi
%---------------------------------------------------------------------
%
\newcount\yearltd\yearltd=\year\advance\yearltd by -1900

\def\Title#1#2{\nopagenumbers\abstractfont\hsize=\hstitle\rightline{#1}%
\vskip 1in\centerline{\titlefont #2}\abstractfont\vskip .5in\pageno=0}
\def\Date#1{\vfill\leftline{#1}\tenpoint\supereject\global\hsize=\hsbody%
\footline={\hss\tenrm\folio\hss}}%      restores pagenumbers
%
%       use following instead of \Date on the preliminary draft,
%       puts date/time on each page in big mode, writes labels in margins

\def\draftmode{\message{ DRAFTMODE }\def\draftdate{{\rm preliminary draft:
\number\month/\number\day/\number\yearltd\ \ \hourmin}}%
\headline={\hfil\draftdate}\writelabels\baselineskip=20pt plus 2pt minus 2pt
 {\count255=\time\divide\count255 by 60 \xdef\hourmin{\number\count255}
  \multiply\count255 by-60\advance\count255 by\time
  \xdef\hourmin{\hourmin:\ifnum\count255<10 0\fi\the\count255}}}
%       use \nolabels to get rid of eqn, ref, and fig labels in draft mode
\def\nolabels{\def\wrlabeL##1{}\def\eqlabeL##1{}\def\reflabeL##1{}}
\def\writelabels{\def\wrlabeL##1{\leavevmode\vadjust{\rlap{\smash%
{\line{{\escapechar=` \hfill\rlap{\sevenrm\hskip.03in\string##1}}}}}}}%
\def\eqlabeL##1{{\escapechar-1\rlap{\sevenrm\hskip.05in\string##1}}}%
\def\reflabeL##1{\noexpand\llap{\noexpand\sevenrm\string\string\string##1}}}
\nolabels
%
% tagged sec numbers
\global\newcount\secno \global\secno=0
\global\newcount\meqno \global\meqno=1
\def\newsec#1{\global\advance\secno by1\message{(\the\secno. #1)}
%\ifx\answ\bigans \vfill\eject \else \bigbreak\bigskip \fi  %if desired
\global\subsecno=0\eqnres@t\noindent{\bf\the\secno. #1}
\writetoca{{\secsym} {#1}}\par\nobreak\medskip\nobreak}
\def\eqnres@t{\xdef\secsym{\the\secno.}\global\meqno=1\bigbreak\bigskip}
\def\sequentialequations{\def\eqnres@t{\bigbreak}}\xdef\secsym{}
\global\newcount\subsecno \global\subsecno=0
\def\subsec#1{\global\advance\subsecno by1\message{(\secsym\the\subsecno.
#1)}
\ifnum\lastpenalty>9000\else\bigbreak\fi
\noindent{\it\secsym\the\subsecno. #1}\writetoca{\string\quad
{\secsym\the\subsecno.} {#1}}\par\nobreak\medskip\nobreak}
\def\appendix#1#2{\global\meqno=1\global\subsecno=0\xdef\secsym{\hbox{#1.}}
\bigbreak\bigskip\noindent{\bf Appendix #1. #2}\message{(#1. #2)}
\writetoca{Appendix {#1.} {#2}}\par\nobreak\medskip\nobreak}
%
%       \eqn\label{a+b=c}       gives displayed equation, numbered
%                               consecutively within sections.
%     \eqnn and \eqna define labels in advance (of eqalign?)
%
\def\eqnn#1{\xdef #1{(\secsym\the\meqno)}\writedef{#1\leftbracket#1}%
\global\advance\meqno by1\wrlabeL#1}
\def\eqna#1{\xdef #1##1{\hbox{$(\secsym\the\meqno##1)$}}
\writedef{#1\numbersign1\leftbracket#1{\numbersign1}}%
\global\advance\meqno by1\wrlabeL{#1$\{\}$}}
\def\eqn#1#2{\xdef #1{(\secsym\the\meqno)}\writedef{#1\leftbracket#1}%
\global\advance\meqno by1$$#2\eqno#1\eqlabeL#1$$}
%
%                            footnotes
\newskip\footskip\footskip14pt plus 1pt minus 1pt %sets footnote baselineskip
\def\footnotefont{\ninepoint}\def\f@t#1{\footnotefont #1\@foot}
\def\f@@t{\baselineskip\footskip\bgroup\footnotefont\aftergroup\@foot\let\next}
\setbox\strutbox=\hbox{\vrule height9.5pt depth4.5pt width0pt}
\global\newcount\ftno \global\ftno=0
\def\foot{\global\advance\ftno by1\footnote{$^{\the\ftno}$}}
%
%say \footend to put footnotes at end
%will cause problems if \ref used inside \foot, instead use \nref before
\newwrite\ftfile
\def\footend{\def\foot{\global\advance\ftno by1\chardef\wfile=\ftfile
$^{\the\ftno}$\ifnum\ftno=1\immediate\openout\ftfile=foots.tmp\fi%
\immediate\write\ftfile{\noexpand\smallskip%
\noexpand\item{f\the\ftno:\ }\pctsign}\findarg}%
\def\footatend{\vfill\eject\immediate\closeout\ftfile{\parindent=20pt
\centerline{\bf Footnotes}\nobreak\bigskip\input foots.tmp }}}
\def\footatend{}
%
%     \ref\label{text}
% generates a number, assigns it to \label, generates an entry.
% To list the refs on a separate page,  \listrefs
%
\global\newcount\refno \global\refno=1
\newwrite\rfile
\def\ref{[\the\refno]\nref}
\def\nref#1{\xdef#1{[\the\refno]}\writedef{#1\leftbracket#1}%
\ifnum\refno=1\immediate\openout\rfile=refs.tmp\fi
\global\advance\refno by1\chardef\wfile=\rfile\immediate
\write\rfile{\noexpand\item{#1\ }\reflabeL{#1\hskip.31in}\pctsign}\findarg}
%        horrible hack to sidestep tex \write limitation
\def\findarg#1#{\begingroup\obeylines\newlinechar=`\^^M\pass@rg}
{\obeylines\gdef\pass@rg#1{\writ@line\relax #1^^M\hbox{}^^M}%
\gdef\writ@line#1^^M{\expandafter\toks0\expandafter{\striprel@x #1}%
\edef\next{\the\toks0}\ifx\next\em@rk\let\next=\endgroup\else\ifx\next\empty%
\else\immediate\write\wfile{\the\toks0}\fi\let\next=\writ@line\fi\next\relax}}
\def\striprel@x#1{} \def\em@rk{\hbox{}}
\def\lref{\begingroup\obeylines\lr@f}
\def\lr@f#1#2{\gdef#1{\ref#1{#2}}\endgroup\unskip}
\def\semi{;\hfil\break}
\def\addref#1{\immediate\write\rfile{\noexpand\item{}#1}} %now unnecessary
\def\startrefs#1{\immediate\openout\rfile=refs.tmp\refno=#1}
\def\xref{\expandafter\xr@f}\def\xr@f[#1]{#1}
\def\refs#1{\count255=1[\r@fs #1{\hbox{}}]}
\def\r@fs#1{\ifx\und@fined#1\message{reflabel \string#1 is undefined.}%
\nref#1{need to supply reference \string#1.}\fi%
\vphantom{\hphantom{#1}}\edef\next{#1}\ifx\next\em@rk\def\next{}%
\else\ifx\next#1\ifodd\count255\relax\xref#1\count255=0\fi%
\else#1\count255=1\fi\let\next=\r@fs\fi\next}
%

%
% this is ugly, but moore insists
\newwrite\ffile\global\newcount\figno \global\figno=1
\def\fig{fig.~\the\figno\nfig}
\def\nfig#1{\xdef#1{fig.~\the\figno}%
\writedef{#1\leftbracket fig.\noexpand~\the\figno}%
\ifnum\figno=1\immediate\openout\ffile=figs.tmp\fi\chardef\wfile=\ffile%
\immediate\write\ffile{\noexpand\medskip\noexpand\item{Fig.\ \the\figno. }
\reflabeL{#1\hskip.55in}\pctsign}\global\advance\figno by1\findarg}
\def\xfig{\expandafter\xf@g}\def\xf@g fig.\penalty\@M\ {}
\def\figs#1{figs.~\f@gs #1{\hbox{}}}
\def\f@gs#1{\edef\next{#1}\ifx\next\em@rk\def\next{}\else
\ifx\next#1\xfig #1\else#1\fi\let\next=\f@gs\fi\next}
\newwrite\lfile
{\escapechar-1\xdef\pctsign{\string\%}\xdef\leftbracket{\string\{}
\xdef\rightbracket{\string\}}\xdef\numbersign{\string\#}}

\def\writestop{\def\writestoppt{\immediate\write\lfile{\string\pageno%
\the\pageno\string\startrefs\leftbracket\the\refno\rightbracket%
\string\def\string\secsym\leftbracket\secsym\rightbracket%
\string\secno\the\secno\string\meqno\the\meqno}\immediate\closeout\lfile}}
\def\writestoppt{}\def\writedef#1{}
\def\seclab#1{\xdef #1{\the\secno}\writedef{#1\leftbracket#1}\wrlabeL{#1=#1}}
\def\subseclab#1{\xdef #1{\secsym\the\subsecno}%
\writedef{#1\leftbracket#1}\wrlabeL{#1=#1}}
\newwrite\tfile \def\writetoca#1{}
\def\leaderfill{\leaders\hbox to 1em{\hss.\hss}\hfill}
%        use this to write file with table of contents
\def\writetoc{\immediate\openout\tfile=toc.tmp
   \def\writetoca##1{{\edef\next{\write\tfile{\noindent ##1
   \string\leaderfill {\noexpand\number\pageno} \par}}\next}}}
%       and this lists table of contents on second pass

%
\catcode`\@=12 % at signs are no longer letters
%
%        Unpleasantness in calling in abstract and title fonts
\edef\tfontsize{\ifx\answ\bigans scaled\magstep3\else scaled\magstep4\fi}
\font\titlerm=cmr10 \tfontsize \font\titlerms=cmr7 \tfontsize
\font\titlermss=cmr5 \tfontsize \font\titlei=cmmi10 \tfontsize
\font\titleis=cmmi7 \tfontsize \font\titleiss=cmmi5 \tfontsize
\font\titlesy=cmsy10 \tfontsize \font\titlesys=cmsy7 \tfontsize
\font\titlesyss=cmsy5 \tfontsize \font\titleit=cmti10 \tfontsize
\skewchar\titlei='177 \skewchar\titleis='177 \skewchar\titleiss='177
\skewchar\titlesy='60 \skewchar\titlesys='60 \skewchar\titlesyss='60
\def\titlefont{\def\rm{\fam0\titlerm}% switch to title font
\textfont0=\titlerm \scriptfont0=\titlerms \scriptscriptfont0=\titlermss
\textfont1=\titlei \scriptfont1=\titleis \scriptscriptfont1=\titleiss
\textfont2=\titlesy \scriptfont2=\titlesys \scriptscriptfont2=\titlesyss
\textfont\itfam=\titleit \def\it{\fam\itfam\titleit}\rm}
 \ifx\answ\bigans\else scaled\magstep1\fi
\ifx\answ\bigans\def\abstractfont{\tenpoint}\else
\font\abssl=cmsl10 scaled \magstep1
\font\absrm=cmr10 scaled\magstep1 \font\absrms=cmr7 scaled\magstep1
\font\absrmss=cmr5 scaled\magstep1 \font\absi=cmmi10 scaled\magstep1
\font\absis=cmmi7 scaled\magstep1 \font\absiss=cmmi5 scaled\magstep1
\font\abssy=cmsy10 scaled\magstep1 \font\abssys=cmsy7 scaled\magstep1
\font\abssyss=cmsy5 scaled\magstep1 \font\absbf=cmbx10 scaled\magstep1
\skewchar\absi='177 \skewchar\absis='177 \skewchar\absiss='177
\skewchar\abssy='60 \skewchar\abssys='60 \skewchar\abssyss='60
\def\abstractfont{\def\rm{\fam0\absrm}% switch to abstract font
\textfont0=\absrm \scriptfont0=\absrms \scriptscriptfont0=\absrmss
\textfont1=\absi \scriptfont1=\absis \scriptscriptfont1=\absiss
\textfont2=\abssy \scriptfont2=\abssys \scriptscriptfont2=\abssyss
\textfont\itfam=\bigit \def\it{\fam\itfam\bigit}\def\footnotefont{\tenpoint}%
\textfont\slfam=\abssl \def\sl{\fam\slfam\abssl}%
\textfont\bffam=\absbf \def\bf{\fam\bffam\absbf}\rm}\fi
\def\tenpoint{\def\rm{\fam0\tenrm}% switch back to 10-point type
\textfont0=\tenrm \scriptfont0=\sevenrm \scriptscriptfont0=\fiverm
\textfont1=\teni  \scriptfont1=\seveni  \scriptscriptfont1=\fivei
\textfont2=\tensy \scriptfont2=\sevensy \scriptscriptfont2=\fivesy
\textfont\itfam=\tenit
\def\it{\fam\itfam\tenit}\def\footnotefont{\ninepoint}%
\textfont\bffam=\tenbf \def\bf{\fam\bffam\tenbf}\def\sl{\fam\slfam\tensl}\rm}
\font\ninerm=cmr9 \font\sixrm=cmr6 \font\ninei=cmmi9 \font\sixi=cmmi6
\font\ninesy=cmsy9 \font\sixsy=cmsy6 \font\ninebf=cmbx9
\font\nineit=cmti9 \font\ninesl=cmsl9 \skewchar\ninei='177
\skewchar\sixi='177 \skewchar\ninesy='60 \skewchar\sixsy='60
\def\ninepoint{\def\rm{\fam0\ninerm}% switch to footnote font
\textfont0=\ninerm \scriptfont0=\sixrm \scriptscriptfont0=\fiverm
\textfont1=\ninei \scriptfont1=\sixi \scriptscriptfont1=\fivei
\textfont2=\ninesy \scriptfont2=\sixsy \scriptscriptfont2=\fivesy
\textfont\itfam=\ninei \def\it{\fam\itfam\nineit}\def\sl{\fam\slfam\ninesl}%
\textfont\bffam=\ninebf \def\bf{\fam\bffam\ninebf}\rm}
%
%---------------------------------------------------------------------
%

\hyphenation{anom-aly anom-alies coun-ter-term coun-ter-terms}
\def\inv{^{\raise.15ex\hbox{${\scriptscriptstyle -}$}\kern-.05em 1}}

\def\Dsl{\,\raise.15ex\hbox{/}\mkern-13.5mu D} %this one can be subscripted
\def\dsl{\raise.15ex\hbox{/}\kern-.57em\partial}

\font\bigit=cmti10 scaled \magstep1
 %pound sterling
\def\lspace{\ifx\answ\bigans{}\else\qquad\fi}
\def\lbspace{\ifx\answ\bigans{}\else\hskip-.2in\fi} % $$\lbspace...$$
\def\boxeqn#1{\vcenter{\vbox{\hrule\hbox{\vrule\kern3pt\vbox{\kern3pt
           \hbox{${\displaystyle #1}$}\kern3pt}\kern3pt\vrule}\hrule}}}
\def\mbox#1#2{\vcenter{\hrule \hbox{\vrule height#2in
               \kern#1in \vrule} \hrule}}  %e.g. \mbox{.1}{.1}
%       matters of taste
%\def\tilde{\widetilde} \def\bar{\overline} \def\hat{\widehat}
%
% some sample definitions
  %     curly letters

\def\e#1{{\rm e}^{^{\textstyle#1}}}

\def\darr#1{\raise1.5ex\hbox{$\leftrightarrow$}\mkern-16.5mu #1}
 %pound sterling

 %puts a small half in a displayed eqn
\def\roughly#1{\raise.3ex\hbox{$#1$\kern-.75em\lower1ex\hbox{$\sim$}}}

%\input harvmac.tex

%%temporary additional macros
% \input macros.tex
% April 16 -- NN

%%%%%%%%%%%%%%%%%%%%%  Rublenye bukvy   %%%%%%%%%%%%%%%%%%%%%%%%
\def\IB{\relax\hbox{$\inbar\kern-.3em{\rm B}$}}
\def\IC{\relax\hbox{$\inbar\kern-.3em{\rm C}$}}
\def\ID{\relax\hbox{$\inbar\kern-.3em{\rm D}$}}
\def\IE{\relax\hbox{$\inbar\kern-.3em{\rm E}$}}
\def\IF{\relax\hbox{$\inbar\kern-.3em{\rm F}$}}
\def\IG{\relax\hbox{$\inbar\kern-.3em{\rm G}$}}
\def\IGa{\relax\hbox{${\rm I}\kern-.18em\Gamma$}}
\def\IH{\relax{\rm I\kern-.18em H}}
\def\IK{\relax{\rm I\kern-.18em K}}
\def\II{\relax{\rm I\kern-.18em I}}
\def\IL{\relax{\rm I\kern-.18em L}}
\def\IP{\relax{\rm I\kern-.18em P}}
\def\IR{\relax{\rm I\kern-.18em R}}
\def\IZ{\relax\ifmmode\mathchoice {\hbox{\cmss Z\kern-.4em Z}}{\hbox{\cmss
Z\kern-.4em Z}} {\lower.9pt\hbox{\cmsss Z\kern-.4em Z}}
{\lower1.2pt\hbox{\cmsss Z\kern-.4em Z}}\else{\cmss Z\kern-.4em Z}\fi}

\def\IB{\relax{\rm I\kern-.18em B}}
\def\IC{{\relax\hbox{$\inbar\kern-.3em{\rm C}$}}}
\def\ID{\relax{\rm I\kern-.18em D}}
\def\IE{\relax{\rm I\kern-.18em E}}
\def\IF{\relax{\rm I\kern-.18em F}}

%%%%%%%%%%%%%%%%%%%% Calligraphic letters  %%%%%%%%%%%%%%%%%%%%%%%

\def\CW {{\cal W}}

%%%%%%%%%%%%%%%%%%%%%%%%%% Derivatives  %%%%%%%%%%%%%%%%%%%%%%%%
\def\p{\partial}

%%Beltrami

%%%%%%%%%%%%%%%%%%%% letters with bar %%%%%%%%%%%%%%%%%%%%%%%%%%

%%%%%%%%%%%%%%%%%%%%%%%%%%% Math symbols %%%%%%%%%%%%%%%%%%%%%%%

%%%%%%%%%%%%%%%%%%%%% Short Cuts %%%%%%%%%%%%%%%%%%%%%%%

\def\demi{{1\over 2}}

%%%%%%%%%%%%%%%%%% Greek %%%%%%%%%%%%%%%%%%%%%%

\def\a{\alpha}
\def\b{\beta}
\def\g{\gamma}  \def\G{\Gamma}
\def\d{\delta}  
\def\m{\mu}
\def\n{\nu}

\def\l{\lambda} 
\def\k{\kappa}
\def\e{\epsilon}

%%%%%%%%%%%%%%%%%% Big ( )  %%%%%%%%%%%%%%%%%%%%%%
\def\|{\Big|}
\def\({\Big(}   \def\){\Big)}
\def\[{\Big[}   \def\]{\Big]}

%%%%%%%%%%%%%%%%%% Text %%%%%%%%%%%%%%%%%%%%%%

%%%%%%%%%%%%% References %%%%%%%%%%%%%%%%%%%%

\def\paper#1#2#3#4{#1, {\sl #2}, #3 {\tt #4}}
% refs with #1=authors, #2=title, #3=publ.ref, #4=hep no :
%\lref\NAME{\paper
%{Authors}{Title(in \it)}{\PLB{No.}{Year}{page},}
%{\hh 0109224 (in\tt)}.}

%\def\hh#1{hep-th/{\it #1}}
\def\hh{hep-th/}

% journal~{\bf no.} (year) page

\def\PLB#1#2#3{Phys. Lett.~{\bf B#1} (#2) #3}
\def\NPB#1#2#3{Nucl. Phys.~{\bf B#1} (#2) #3}
\def\PRL#1#2#3{Phys. Rev. Lett.~{\bf #1} (#2) #3}
\def\CMP#1#2#3{Comm. Math. Phys.~{\bf #1} (#2) #3}
\def\PRD#1#2#3{Phys. Rev.~{\bf D#1} (#2) #3}
\def\MPL#1#2#3{Mod. Phys. Lett.~{\bf #1} (#2) #3}
\def\IJMP#1#2#3{Int. Jour. Mod. Phys.~{\bf #1} (#2) #3}

%%%%%%%%%%%%%%%%%%% Something to deal with sub-sub-sections
%%%%%%%%%%%%%%%%%%%%%%%%%%%%%%%%%%%%%%%%%%%%%%%

\def\unlockat{\catcode`\@=11}
\def\lockat{\catcode`\@=12}

\unlockat

% Something to deal with sub-sub-sections

\def\newsec#1{\global\advance\secno by1\message{(\the\secno. #1)}
\global\subsecno=0\global\subsubsecno=0\eqnres@t\noindent {\bf\the\secno. #1}
\writetoca{{\secsym} {#1}}\par\nobreak\medskip\nobreak}
\global\newcount\subsecno \global\subsecno=0
\def\subsec#1{\global\advance\subsecno by1\message{(\secsym\the\subsecno.
#1)}
\ifnum\lastpenalty>9000\else\bigbreak\fi\global\subsubsecno=0
\noindent{\it\secsym\the\subsecno. #1}
\writetoca{\string\quad {\secsym\the\subsecno.} {#1}}
\par\nobreak\medskip\nobreak}
\global\newcount\subsubsecno \global\subsubsecno=0
\def\subsubsec#1{\global\advance\subsubsecno by1
\message{(\secsym\the\subsecno.\the\subsubsecno. #1)}
\ifnum\lastpenalty>9000\else\bigbreak\fi
\noindent\quad{\secsym\the\subsecno.\the\subsubsecno.}{#1}
\writetoca{\string\qquad{\secsym\the\subsecno.\the\subsubsecno.}{#1}}
\par\nobreak\medskip\nobreak}

\def\subsubseclab#1{\DefWarn#1\xdef #1{\noexpand\hyperref{}{subsubsection}%
{\secsym\the\subsecno.\the\subsubsecno}%
{\secsym\the\subsecno.\the\subsubsecno}}%
\writedef{#1\leftbracket#1}\wrlabeL{#1=#1}}% Macros for boxes
\lockat

%why???\font\manual=manfnt
\def\dbend{\lower3.5pt\hbox{\manual\char127}}

%%%%%%%%%%%%%%%%%%% Macros for boxes %%%%%%%%%%%%%%%%%%

\def\boxit#1{\vbox{\hrule\hbox{\vrule\kern8pt
\vbox{\hbox{\kern8pt}\hbox{\vbox{#1}}\hbox{\kern8pt}}
\kern8pt\vrule}\hrule}}

\def\mathboxit#1{\vbox{\hrule\hbox{\vrule\kern8pt\vbox{\kern8pt
\hbox{$\displaystyle #1$}\kern8pt}\kern8pt\vrule}\hrule}}

%%%%%%%%%%%%%%%%%%%% ANOTHER SET OF MACROS %%%%%%%%%%%%%%%%%%

\def\inbar{\,\vrule height1.5ex width.4pt depth0pt}

\font\cmss=cmss10 \font\cmsss=cmss10 at 7pt

%REFERENCES
%%%%%%%%%%%%%%%%%%%%%%%%%%%%%%%%%%%%%%%%%%%%%%%%%%%

\lref\simons{ J. Cheeger and J. Simons, {\it Differential Characters and
Geometric Invariants},  Stony Brook Preprint, (1973), unpublished.}

\lref\cargese{ L.~Baulieu, {\it Algebraic quantization of gauge theories},
Perspectives in fields and particles, Plenum Press, eds. Basdevant-Levy,
Cargese Lectures 1983}

\lref\antifields{ L. Baulieu, M. Bellon, S. Ouvry, C.Wallet, Phys.Letters
B252 (1990) 387; M.  Bocchichio, Phys. Lett. B187 (1987) 322;  Phys. Lett. B
192 (1987) 31; R.  Thorn    Nucl. Phys.   B257 (1987) 61. }

\lref\thompson{ George Thompson,  Annals Phys. 205 (1991) 130; J.M.F.
Labastida, M. Pernici, Phys. Lett. 212B  (1988) 56; D. Birmingham, M.Blau,
M. Rakowski and G.Thompson, Phys. Rept. 209 (1991) 129.}

\lref\tonin{ Tonin}

\lref\wittensix{ E.  Witten, {\it New  Gauge  Theories In Six Dimensions},
\hh{9710065}. }

\lref\orlando{ O. Alvarez, L. A. Ferreira and J. Sanchez Guillen, {\it  A New
Approach to Integrable Theories in any Dimension}, hep-th/9710147.}

\lref\wittentopo{ E.  Witten,  {\it  Topological Quantum Field Theory},
\hh9403195, Commun.  Math. Phys.  {117} (1988)353.  }

\lref\wittentwist{ E.  Witten, {\it Supersymmetric Yang--Mills theory on a
four-manifold}, J.  Math.  Phys.  {35} (1994) 5101.}

\lref\west{ L.~Baulieu, P.~West, {\it Six Dimensional TQFTs and  Self-dual
Two-Forms,} Phys.Lett. B {\bf 436 } (1998) 97, /hep-th/9805200}

\lref\bv{ I.A. Batalin and V.A. Vilkowisky,    Phys. Rev.   D28  (1983)
2567\semi M. Henneaux,  Phys. Rep.  126   (1985) 1\semi M. Henneaux and C.
Teitelboim, {\it Quantization of Gauge Systems}
  Princeton University Press,  Princeton (1992).}

\lref\kyoto{ L. Baulieu, E. Bergschoeff and E. Sezgin, Nucl. Phys.
B307(1988)348\semi L. Baulieu,   {\it Field Antifield Duality, p-Form Gauge
Fields
   and Topological Quantum Field Theories}, hep-th/9512026,
   Nucl. Phys. B478 (1996) 431.  }

\lref\sourlas{ G. Parisi and N. Sourlas, {\it Random Magnetic Fields,
Supersymmetry and Negative Dimensions}, Phys. Rev. Lett.  43 (1979) 744;
Nucl.  Phys.  B206 (1982) 321.  }

\lref\SalamSezgin{ A.  Salam  and  E.  Sezgin, {\it Supergravities in
diverse dimensions}, vol.  1, p. 119\semi P.  Howe, G.  Sierra and P.
Townsend, Nucl Phys B221 (1983) 331.}

\lref\nekrasov{ A. Losev, G. Moore, N. Nekrasov, S. Shatashvili, {\it
Four-Dimensional Avatars of Two-Dimensional RCFT},  hep-th/9509151, Nucl.
Phys.  Proc.  Suppl.   46 (1996) 130\semi L.  Baulieu, A.  Losev,
N.~Nekrasov  {\it Chern-Simons and Twisted Supersymmetry in Higher
Dimensions},  hep-th/9707174, to appear in Nucl.  Phys.  B.  }

\lref\WitDonagi{R.~ Donagi, E.~ Witten, ``Supersymmetric Yang--Mills Theory
and Integrable Systems'', hep-th/9510101, Nucl. Phys.{\bf B}460 (1996)
299-334}
\lref\Witfeb{E.~ Witten, ``Supersymmetric Yang--Mills Theory On A
Four-Manifold,''  hep-th/9403195; J. Math. Phys. {\bf 35} (1994) 5101.}
\lref\Witgrav{E.~ Witten, ``Topological Gravity'', Phys.Lett.206B:601, 1988}
\lref\witaffl{I. ~ Affleck, J.A.~ Harvey and E.~ Witten,
        ``Instantons and (Super)Symmetry Breaking
        in $2+1$ Dimensions'', Nucl. Phys. {\bf B}206 (1982) 413}
\lref\wittabl{E.~ Witten,  ``On $S$-Duality in Abelian Gauge Theory,''
hep-th/9505186; Selecta Mathematica {\bf 1} (1995) 383}
\lref\wittgr{E.~ Witten, ``The Verlinde Algebra And The Cohomology Of The
Grassmannian'',  hep-th/9312104}
\lref\wittenwzw{E. Witten, ``Non abelian bosonization in two dimensions,''
Commun. Math. Phys. {\bf 92} (1984)455 }
\lref\witgrsm{E. Witten, ``Quantum field theory, grassmannians and algebraic
curves,'' Commun.Math.Phys.113:529,1988}
\lref\wittjones{E. Witten, ``Quantum field theory and the Jones
polynomial,'' Commun.  Math. Phys., 121 (1989) 351. }
\lref\witttft{E.~ Witten, ``Topological Quantum Field Theory", Commun. Math.
Phys. {\bf 117} (1988) 353.}
\lref\wittmon{E.~ Witten, ``Monopoles and Four-Manifolds'', hep-th/9411102}
\lref\Witdgt{ E.~ Witten, ``On Quantum gauge theories in two dimensions,''
Commun. Math. Phys. {\bf  141}  (1991) 153}
\lref\witrevis{E.~ Witten,
 ``Two-dimensional gauge theories revisited'', hep-th/9204083; J. Geom.
Phys. 9 (1992) 303-368}
\lref\Witgenus{E.~ Witten, ``Elliptic Genera and Quantum Field Theory'',
Comm. Math. Phys. 109(1987) 525. }
\lref\OldZT{E. Witten, ``New Issues in Manifolds of SU(3) Holonomy,'' {\it
Nucl. Phys.} {\bf B268} (1986) 79 \semi J. Distler and B. Greene, ``Aspects
of (2,0) String Compactifications,'' {\it Nucl. Phys.} {\bf B304} (1988) 1
\semi B. Greene, ``Superconformal Compactifications in Weighted Projective
Space,'' {\it Comm. Math. Phys.} {\bf 130} (1990) 335.}
\lref\bagger{E.~ Witten and J. Bagger, Phys. Lett. {\bf 115B}(1982) 202}
\lref\witcurrent{E.~ Witten,``Global Aspects of Current Algebra'',
Nucl.Phys.B223 (1983) 422\semi ``Current Algebra, Baryons and Quark
Confinement'', Nucl.Phys. B223 (1993) 433}
\lref\Wittreiman{S.B. Treiman, E. Witten, R. Jackiw, B. Zumino, ``Current
Algebra and Anomalies'', Singapore, Singapore: World Scientific ( 1985) }
\lref\Witgravanom{L. Alvarez-Gaume, E.~ Witten, ``Gravitational Anomalies'',
Nucl.Phys.B234:269,1984. }

\lref\nicolai{\paper {H.~Nicolai}{New Linear Systems for 2D Poincar\'e
Supergravities}{\NPB{414}{1994}{299},}{\hh 9309052}.}

%%%%%%
%% References herein
%%%%%%%%%%%%%%%%%%%%%%%%%%%%%%%%%%%%%%%%%%%%%%%%

%%\lref\NAME{\paper
%%{Authors}{Title(in \sl)}{\PLB{No.}{Year}{page},}
%%{\hh 0109224 (in\tt)}.}

\lref\baex{\paper {L.~Baulieu, B.~Grossman}{Monopoles and Topological Field
Theory}{\PLB{214}{1988}{223}.}{}\paper {L.~Baulieu}{Chern-Simons
Three-Dimensional and
Yang--Mills-Higgs Two-Dimensional Systems as Four-Dimensional Topological
Quantum Field Theories}{\PLB{232}{1989}{473}.}{}}

\lref\bg{\paper {L.~Baulieu, B.~Grossman}{Monopoles and Topological Field
Theory}{\PLB{214}{1988}{223}.}{}}

\lref\seibergsix{\paper {N.~Seiberg}{Non-trivial Fixed Points of The
Renormalization Group in Six
 Dimensions}{\PLB{390}{1997}{169}}{\hh 9609161}\semi
\paper {O.J.~Ganor, D.R.~Morrison, N.~Seiberg}{
  Branes, Calabi-Yau Spaces, and Toroidal Compactification of the N=1
  Six-Dimensional $E_8$ Theory}{\NPB{487}{1997}{93}}{\hh 9610251}\semi
\paper {O.~Aharony, M.~Berkooz, N.~Seiberg}{Light-Cone
  Description of (2,0) Superconformal Theories in Six
  Dimensions}{Adv. Theor. Math. Phys. {\bf 2} (1998) 119}{\hh 9712117.}}

\lref\cs{\paper {L.~Baulieu}{Chern-Simons Three-Dimensional and
Yang--Mills-Higgs Two-Dimensional Systems as Four-Dimensional Topological
Quantum Field Theories}{\PLB{232}{1989}{473}.}{}}

\lref\beltrami{\paper {L.~Baulieu, M.~Bellon}{Beltrami Parametrization and
String Theory}{\PLB{196}{1987}{142}}{}\semi
\paper {L.~Baulieu, M.~Bellon, R.~Grimm}{Beltrami Parametrization For
Superstrings}{\PLB{198}{1987}{343}}{}\semi
\paper {R.~Grimm}{Left-Right Decomposition of Two-Dimensional Superspace
Geometry and Its BRS Structure}{Annals Phys. {\bf 200} (1990) 49.}{}}

\lref\bbg{\paper {L.~Baulieu, M.~Bellon, R.~Grimm}{Left-Right Asymmetric
Conformal Anomalies}{\PLB{228}{1989}{325}.}{}}

\lref\bonora{\paper {G.~Bonelli, L.~Bonora, F.~Nesti}{String Interactions
from Matrix String Theory}{\NPB{538}{1999}{100},}{\hh 9807232}\semi
\paper {G.~Bonelli, L.~Bonora, F.~Nesti, A.~Tomasiello}{Matrix String Theory
and its Moduli Space}{}{\hh 9901093.}}

\lref\corrigan{\paper {E.~Corrigan, C.~Devchand, D.B.~Fairlie,
J.~Nuyts}{First Order Equations for Gauge Fields in Spaces of Dimension
Greater Than Four}{\NPB{214}{452}{1983}.}{}}

\lref\acha{\paper {B.S.~Acharya, M.~O'Loughlin, B.~Spence}{Higher
Dimensional Analogues of Donaldson-Witten Theory}{\NPB{503}{1997}{657},}{\hh
9705138}\semi
\paper {B.S.~Acharya, J.M.~Figueroa-O'Farrill, M.~O'Loughlin,
B.~Spence}{Euclidean
  D-branes and Higher-Dimensional Gauge   Theory}{\NPB{514}{1998}{583},}{\hh
  9707118.}}

\lref\Witr{\paper{E.~Witten}{Introduction to Cohomological Field   Theories}
{Lectures at Workshop on Topological Methods in Physics (Trieste, Italy, Jun
11-25, 1990), \IJMP{A6}{1991}{2775}.}{}}

\lref\ohta{\paper {L.~Baulieu, N.~Ohta}{Worldsheets with Extended
Supersymmetry} {\PLB{391}{1997}{295},}{\hh 9609207}.}

\lref\gravity{\paper {L.~Baulieu}{Transmutation of Pure 2-D Supergravity
Into Topological 2-D Gravity and Other Conformal Theories}
{\PLB{288}{1992}{59},}{\hh 9206019.}}

\lref\wgravity{\paper {L.~Baulieu, M.~Bellon, R.~Grimm}{Some Remarks on  the
Gauging of the Virasoro and   $w_{1+\infty}$
Algebras}{\PLB{260}{1991}{63}.}{}}

\lref\fourd{\paper {E.~Witten}{Topological Quantum Field
Theory}{\CMP{117}{1988}{353}}{}\semi
\paper {L.~Baulieu, I.M.~Singer}{Topological Yang--Mills Symmetry}{Nucl.
Phys. Proc. Suppl. {\bf 15B} (1988) 12.}{}}

\lref\topo{\paper {L.~Baulieu}{On the Symmetries of Topological Quantum Field
Theories}{\IJMP{A10}{1995}{4483},}{\hh 9504015}\semi
\paper {R.~Dijkgraaf, G.~Moore}{Balanced Topological Field
Theories}{\CMP{185}{1997}{411},}{\hh 9608169.}}

\lref\wwgravity{\paper {I.~Bakas} {The Large $N$ Limit   of Extended
Conformal Symmetries}{\PLB{228}{1989}{57}.}{}}

\lref\wwwgravity{\paper {C.M.~Hull}{Lectures on $\CW$-Gravity,
$\CW$-Geometry and
$\CW$-Strings}{}{\hh 9302110}, and~references therein.}

\lref\wvgravity{\paper {A.~Bilal, V.~Fock, I.~Kogan}{On the origin of
$W$-algebras}{\NPB{359}{1991}{635}.}{}}

\lref\surprises{\paper {E.~Witten} {Surprises with Topological Field
Theories} {Lectures given at ``Strings 90'', Texas A\&M, 1990,}{Preprint
IASSNS-HEP-90/37.}}

\lref\stringsone{\paper {L.~Baulieu, M.B.~Green, E.~Rabinovici}{A Unifying
Topological Action for Heterotic and  Type II Superstring  Theories}
{\PLB{386}{1996}{91},}{\hh 9606080.}}

\lref\stringsN{\paper {L.~Baulieu, M.B.~Green, E.~Rabinovici}{Superstrings
from   Theories with $N>1$ World Sheet Supersymmetry}
{\NPB{498}{1997}{119},}{\hh 9611136.}}

\lref\bks{\paper {L.~Baulieu, H.~Kanno, I.~Singer}{Special Quantum Field
Theories in Eight and Other Dimensions}{\CMP{194}{1998}{149},}{\hh
9704167}\semi
\paper {L.~Baulieu, H.~Kanno, I.~Singer}{Cohomological Yang--Mills Theory
  in Eight Dimensions}{ Talk given at APCTP Winter School on Dualities in
String Theory (Sokcho, Korea, February 24-28, 1997),} {\hh 9705127.}}

\lref\witdyn{\paper {P.~Townsend}{The eleven dimensional supermembrane
revisited}{\PLB{350}{1995}{184},}{\hh9501068}\semi
\paper{E.~Witten}{String Theory Dynamics in Various Dimensions}
{\NPB{443}{1995}{85},}{\hh 9503124}.}

\lref\bfss{\paper {T.~Banks, W.Fischler, S.H.~Shenker,
L.~Susskind}{$M$-Theory as a Matrix Model~:
A~Conjecture}{\PRD{55}{1997}{5112},} {\hh9610043.}}

\lref\seiberg{\paper {N.~Seiberg}{Why is the Matrix Model
Correct?}{\PRL{79}{1997}{3577},} {\hh 9710009.}}

\lref\sen{\paper {A.~Sen}{$D0$ Branes on $T^n$ and Matrix Theory}{Adv.
Theor. Math. Phys.~{\bf 2} (1998) 51,} {\hh 9709220.}}

\lref\laroche{\paper {L.~Baulieu, C.~Laroche} {On Generalized Self-Duality
Equations Towards Supersymmetric   Quantum Field Theories Of
Forms}{\MPL{A13}{1998}{1115},}{\hh  9801014.}}

\lref\bsv{\paper {M.~Bershadsky, V.~Sadov, C.~Vafa} {$D$-Branes and
Topological Field Theories}{\NPB{463} {1996}{420},}{\hh9511222.}}

\lref\vafapuzz{\paper {C.~Vafa}{Puzzles at Large N}{}{\hh 9804172.}}

\lref\dvv{\paper {R.~Dijkgraaf, E.~Verlinde, H.~Verlinde} {Matrix String
Theory}{\NPB{500}{1997}{43},} {\hh9703030.}}

\lref\wynter{\paper {T.~Wynter}{Gauge Fields and Interactions in Matrix
String Theory}{\PLB{415}{1997}{349},}{\hh9709029.}}

\lref\kvh{\paper {I.~Kostov, P.~Vanhove}{Matrix String Partition
Functions}{}{\hh9809130.}}

\lref\ikkt{\paper {N.~Ishibashi, H.~Kawai, Y.~Kitazawa, A.~Tsuchiya} {A
Large $N$ Reduced Model as Superstring}{\NPB{498} {1997}{467},}{\hh
9612115.}}

\lref\ss{\paper {S.~Sethi, M.~Stern} {$D$-Brane Bound States
Redux}{\CMP{194}{1998} {675},}{\hh 9705046.}}

\lref\mns{\paper {G.~Moore, N.~Nekrasov, S.~Shatashvili} {$D$-particle Bound
States and Generalized Instantons}{} {\hh 9803265.}}

\lref\bsh{\paper {L.~Baulieu, S.~Shatashvili} {Duality from Topological
Symmetry}{} {\hh 9811198.}}

\lref\pawu{ {G.~Parisi, Y.S.~Wu,} {}{ Sci. Sinica  {\bf 24} {(1981)} {484}.}}

%%%%%%%%%%%%%%%%
\lref\lbpert{ {L.~Baulieu,}   {\it Pertrubative gauge theories}, {Physics
Reports {\bf 129 } (1985) 1.} {}}

\lref\hirschfeld{ {P.~Hirschfeld,}   {\it } {Nucl. Phys.
{\bf 157} (1979) 37.} {}}

\lref\fleepr{ {R.~Friedberg, T.~D.~Lee, Y.~Pang, H.~C.~Ren,}   {\it A soluble
gauge model with Gribov-type copies}, {Ann. of Phys. {\bf 246 } (1996) 381.} {}}

\lref\coulomb{ {L.~Baulieu, D.~Zwanziger, }   {\it Renormalizable
Non-Covariant Gauges and Coulomb Gauge Limit}, {Nucl. Phys. B {\bf 548 }
(1999) 527-562.} {\hh 9807024}.}

\lref\coulham{ {D.~Zwanziger, }   {\it Lattice Coulomb hamiltonian and static
color-Coulomb field}, {Nucl. Phys. B {\bf 485 } (1997) 185-240.} {}}

\lref\rcoulomb{ {D.~Zwanziger, }   {\it Renormalization in the Coulomb
gauge and order parameter for confinement in QCD}, {Nucl.Phys. B {\bf 518
} (1998) 237-272.} {}}

\lref\kogsuss{ {J.~Kogut and L. Susskind, }   {\it } {Phys. Rev. D {\bf 11}
(1975) 395.} {}}

\lref\kugoojima{ {T.~Kugo and I. Ojima, }   {\it Local covariant
operator formalism of non-Abelian gauge theories and quark confinement
problem}, {Suppl. Prog. Theor. Phys. {\bf 66 } (1979) 1-130.} {}}

\lref\horne{ {J.H.~Horne, }   {\it
Superspace versions of Topological Theories}, {Nucl.Phys. B {\bf 318
} (1989) 22.} {}}

\lref\sto{ {S.~Ouvry, R.~Stora, P.~Van~Baal }   {\it
}, {Phys. Lett. B {\bf 220
} (1989) 159;} {}{ R.~Stora, {\it Exercises in   Equivariant Cohomology},
In Quabtum Fields and Quantum Space Time, Edited
by 't Hooft et al., Plenum Press, New York, 1997}            }

\lref\dan{ {D.~Zwanziger},  {\it Covariant Quantization of Gauge
Fields without Gribov Ambiguity}, {Nucl. Phys. B {\bf   192}, (1981)
{259}.}{}}

\lref\danlau{ {L.~Baulieu, D.~Zwanziger, } {\it Equivalence of Stochastic
Quantization and the-Popov Ansatz,
  }{Nucl. Phys. B  {\bf 193 } (1981) {163}.}{}}

\lref\dzgribreg{ {D.~Zwanziger},  {\it Non-perturbative modification of the
Faddeev-Popov formula and banishment of the naive vacuum}, {Nucl. Phys. B
{\bf   209}, (1982) {336}.}{}}

\lref\danzinn{  {J.~Zinn-Justin, D.~Zwanziger, } {}{Nucl. Phys. B  {\bf
295} (1988) {297}.}{}}

\lref\dzvan{ {D.~Zwanziger, }   {\it Vanishing of zero-momentum lattice
gluon propagator and color confinement}, {Nucl.Phys. B {\bf 364 }
(1991) 127.} }

\lref\critical{ {D.~Zwanziger},  {\it Critical limit of lattice gauge
theory}, {Nucl. Phys. B {\bf 378} (1992) {525-590}.}}

\lref\horizcon{ {D.~Zwanziger, }   {\it Renormalizability of the critical limit
of lattice gauge theory by BRS invariance,} {Nucl. Phys. B {\bf 399 } (1993)
477.} }

\lref\horizcona{ {D.~Zwanziger, }   {\it Fundamental modular region, Boltzmann
factor and area law in lattice theory }, {Nucl.Phys. B {\bf 412 }
(1994) 657.} }

\lref\horizpt{ {M.~Schaden and D.~Zwanziger, }   
{\it Horizon condition holds pointwise on finite lattice with free boundary
condition,} {hep-th/9410019.} }

\lref\czcoulf{ {A.~Cucchieri and D.~Zwanziger, }   {\it Static color-Coulomb
force}, {Phys. Rev. Lett. {\bf 78 } (1997) 3814} }

\lref\acdland{ {A.~Cucchieri,}   {\it Infrared behavior of the gluon
propagator in lattice landau gauge: the three-dimensional case,} 
{Phys. Rev. {\bf D60}:034508, 1999}}

\lref\cznumstgl{ {A.~Cucchieri, and D.~Zwanziger}   {\it Numerical study of
gluon propagator and confinement scenario in minimal Coulomb gauge}, 
{Phys. Rev. {\bf D65}:014001, 2002}}

\lref\czfitgrib{ {A.~Cucchieri, and D.~Zwanziger}   {\it Fit to gluon
propagator and Gribov formula}, {Phys. Lett. {\bf B524}:123-128, 2002}}

\lref\suman{ {H.~Suman, and K.~Schilling}   {\it } 
Phys. Lett. {\bf B373} (1996) 314.}

\lref\bonnet{ {F.~Bonnet, P.~O.~Bowman, D.~B.~Leinweber, and A.~G.~Williams}  
{\it }  Phys. Rev. {\bf D62} (2000) 051501.}

\lref\nakamuraa{ {A. Nakamura and M. Mizutani}   {\it Numerical study of
gauge fixing ambiguity,}  Vistas in Astronomy {\bf 37} (1993)
305.}

\lref\nakamurab{ {M. Mizutani and A. Nakamura}   {\it Stochastic gauge
fixing for compact lattice gauge theories,}  Nucl. Phys. B (Proc. Suppl.)
{\bf 34} (1994) 253.}

\lref\nakamurac{ {H. Aiso, M. Fukuda, T. Iwamiya, A. Nakamura, T. Nakamura,
and M. Yoshida }   {\it Gauge fixing and gluon propagators,}  Prog. Theor.
Physics. (Suppl.) {\bf 122} (1996) 123.}

\lref\nakamurad{ {H. Aiso, J. Fromm, M. Fukuda, T. Iwamiya, A. Nakamura, T.
Nakamura, M. Stingl and M. Yoshida }   {\it Towards understanding of
confinement of gluons,}  Nucl. Phys. B (Proc. Suppl.) {\bf 53} (1997) 570.}

\lref\nakamurae{ {F.~Shoji, T.~Suzuki, H.~Kodama, and A.~Nakamura,}  
{\it }  Phys. Lett. {\bf B476} (2000) 199.}

\lref\direnzo{ {F.~DiRenzo, L.~Scorzato,}  
{\it Lattice 99,}  Nucl. Phys. B (Proc. Suppl.) {\bf 83-84} (2000) 822.}

\lref\munoz{ { A.~Munoz Sudupe, R. F. Alvarez-Estrada, } {}
Phys. Lett. {\bf 164} (1985) 102; {} {\bf 166B} (1986) 186. }

\lref\okano{ { K.~Okano, } {}
Nucl. Phys. {\bf B289} (1987) 109; {} Prog. Theor. Phys.
suppl. {\bf 111} (1993) 203. }

\lref\baugros{ {L.~Baulieu, B.~Grossman, } {\it A topological Interpretation
of  Stochastic Quantization} {Phys. Lett. B {\bf  212} {(1988)} {351}.}}

\lref\bautop{ {L.~Baulieu}{ \it Stochastic and Topological Field Theories},
{Phys. Lett. B {\bf   232} (1989) {479}}{}; {}{ \it Topological Field Theories
And Gauge Invariance in Stochastic Quantization}, {Int. Jour. Mod.  Phys. A
{\bf  6} (1991) {2793}.}{}}

\lref\bautopr{  {L.~Baulieu, B.~Grossman, } {\it A topological Interpretation
of  Stochastic Quantization} {Phys. Lett. B {\bf  212} {(1988)} {351}};
 {L.~Baulieu}{ \it Stochastic and Topological Field Theories},
{Phys. Lett. B {\bf   232} (1989) {479}}{}; {}{ \it Topological Field Theories
And Gauge Invariance in Stochastic Quantization}, {Int. Jour. Mod.  Phys. A
{\bf  6} (1991) {2793}.}{}}

\lref\bautoprr{  {L.~Baulieu, B.~Grossman, } { } {Phys. Lett. B {\bf  212}
{(1988)} {351}};
 {L.~Baulieu}{ },
{Phys. Lett. B {\bf   232} (1989) {479}}{}; {}{  }, {Int. Jour. Mod.
Phys. A {\bf  6} (1991) {2793}.}{}}
\lref\samson{ {L.~Baulieu, S.L.~Shatashvili, { \it Duality from Topological
Symmetry}, {JHEP {\bf 9903} (1999) 011, hep-th/9811198.}}}{}

\lref\halpern{ {H.S.~Chan, M.B.~Halpern}{}, {Phys. Rev. D {\bf   33} (1986)
{540}.}}

\lref\yue{ {Yue-Yu}, {Phys. Rev. D {\bf   33} (1989) {540}.}}

\lref\neuberger{ {H.~Neuberger,} {\it Non-perturbative gauge Invariance},
{ Phys. Lett. B {\bf 175} (1986) {69}.}{}}

\lref\gribov{  {V.N.~Gribov,} {}{Nucl. Phys. B {\bf 139} (1978) {1}.}{}}

\lref\huffel{ {P.H.~Daamgard, H. Huffel},  {}{Phys. Rep. {\bf 152} (1987)
{227}.}{}}

\lref\creutz{ {M.~Creutz},  {\it Quarks, Gluons and  Lattices, }  Cambridge
University Press 1983, pp 101-107.}

\lref\zinn{ {J. ~Zinn-Justin, }  {Nucl. Phys. B {\bf  275} (1986) {135}.}}

\lref\gozzi{ {E. ~Gozzi,} {\it Functional Integral approach to Parisi--Wu
Quantization: Scalar Theory,} { Phys. Rev. {\bf D28} (1983) {1922}.}}

\lref\singer{
 I.M. Singer, { Comm. of Math. Phys. {\bf 60} (1978) 7.}}

\lref\neu{ {H.~Neuberger,} {Phys. Lett. B {\bf 295}
(1987) {337}.}{}}

\lref\testa{ {M.~Testa,} {}{Phys. Lett. B {\bf 429}
(1998) {349}.}{}}

\lref\Martin{ L.~Baulieu and M. Schaden, {\it Gauge Group TQFT and Improved
Perturbative Yang--Mills Theory}, {  Int. Jour. Mod.  Phys. A {\bf  13}
(1998) 985},   hep-th/9601039.}

\lref\ostseil { K.~Osterwalder and E.~Seiler, {\it Gauge field theories on
the lattice} {  Ann. Phys. {\bf 110} (1978) 440}.}

\lref\fradshen { E.~Fradken and S.~Shenker, {\it Phase diagrams of lattice
guage theories with Higgs fields} {  Phys. Rev. {\bf D19} (1979) 3682}.}

\lref\banksrab{ T.~Banks and E.~Rabinovici, {} {  Nucl. Phys. {\bf B160}
(1979) 349}.}

\lref\nielsen{N. K. Nielsen, {\it On The Gauge Dependence Of Spontaneous
Symmetry Breaking In Gauge Theories},
Nucl.\ Phys.\ {\bf B101}, 173 (1975)}

\lref\nadkarni{ S. Nadkarni, {\it The SU(2) adjoint Higgs model in three
dimensions } {  Nucl. Phys. {\bf B334} (1990) 559}.}

\lref\stackteper{ A.~Hart, O.~Philipsen, J.~D.~Stack, and M.~Teper,
{\it On the phase diagram of the SU(2) adjoint Higgs model in 2+1
dimensions } { hep-lat/9612021}.}

\lref\kajantie{K.~Kajantie, M.~Laine, K.~Rummujkainen, M.~Shaposhnikov
{\it 3D SU(N)+adjoint Higgs theory and finite temperature QCD }
{hep-ph/9704416}.}

\lref\batrouni{G. G. Batrouni, G. R. Katz, A. S. Kronfeld, G. P. Lepage,
B. Svetitsky and K. G. Wilson, {} {Phys. Rev. {\bf D32} (1985) 2736}}

\lref\davies{C. T. H. Davies, G. G. Batrouni, G. R. Katz, A. S. Kronfeld,
G. P. Lepage, K. G. Wilson, P. Rossi and B. Svetitsky, {} {Phys. Rev. {\bf
D41} (1990) 1953}}

\lref\fukugita{M. Fukugita, Y. Oyanagi and A. Ukawa, {}
{Phys. Rev. Lett. {\bf 57} (1986) 953;
Phys. Rev. {\bf D36} (1987) 824}}

\lref\kronfeld{A. S. Kronfeld, {\it Dynamics of Langevin Simulation} {Prog.
Theor. Phys. Suppl. {\bf 111} (1993) 293}}

\lref\polyakov{ A.~Polyakov, {} {Phys. Letts. {\bf B59} (1975) 82;
Nucl. Phys. {\bf B120} (1977) 429}; {\it Gauge fields and strings,}
ch. 4 (Harwood Academic Publishers, 1987).}

\lref\thooft{ G.~'t Hooft, {} {Nucl. Phys. {\bf B79} (1974) 276}; {}
Nucl. Phys. {\bf B190} (1981) 455 {}.}

\lref\elitzur{S.~Elitzur, {} {Phys. Rev. {\bf D12} (1975) 3978}}

%\polyakov \nadkarni \stackteper \kajantie

%%%%%%%%%%%%%%%%%%%%%%%%%%%%%%%%%%%%%%%%%%%%%%%%%%%%%%%%%%%%%%%%%
\lref\baugros{ {L.~Baulieu, B.~Grossman, } {\it A topological Interpretation
of  Stochastic Quantization} {Phys. Lett. B {\bf  212} {(1988)} {351}.}}

\lref\bautop{ {L.~Baulieu}{ \it Stochastic and Topological Field Theories},
{Phys. Lett. B {\bf   232} (1989) {479}}{}; {}{ \it Topological Field Theories
And Gauge Invariance in Stochastic Quantization}, {Int. Jour. Mod.  Phys. A
{\bf  6} (1991) {2793}.}{}}

\lref\bautopr{  {L.~Baulieu, B.~Grossman, } {\it A topological Interpretation
of  Stochastic Quantization} {Phys. Lett. B {\bf  212} {(1988)} {351}};
 {L.~Baulieu}{ \it Stochastic and Topological Field Theories},
{Phys. Lett. B {\bf   232} (1989) {479}}{}; {}{ \it Topological Field Theories
And Gauge Invariance in Stochastic Quantization}, {Int. Jour. Mod.  Phys. A
{\bf  6} (1991) {2793}.}{}}

\lref\samson{ {L.~Baulieu, S.L.~Shatashvili, { \it Duality from Topological
Symmetry}, {JHEP {\bf 9903} (1999) 011, hep-th/9811198.}}}{}

\lref\halpern{ {H.S.~Chan, M.B.~Halpern}{}, {Phys. Rev. D {\bf   33} (1986)
{540}.}}

\lref\yue{ {Yue-Yu}, {Phys. Rev. D {\bf   33} (1989) {540}.}}

\lref\neuberger{ {H.~Neuberger,} {\it Non-perturbative gauge Invariance},
{ Phys. Lett. B {\bf 175} (1986) {69}.}{}}

\lref\huffel{ {P.H.~Daamgard, H. Huffel},  {}{Phys. Rep. {\bf 152} (1987)
{227}.}{}}

\lref\creutz{ {M.~Creutz},  {\it Quarks, Gluons and  Lattices, }  Cambridge
University Press 1983, pp 101-107.}

\lref\zinn{ {J. ~Zinn-Justin, }  {Nucl. Phys. B {\bf  275} (1986) {135}.}}

\lref\shamir{  {Y.~Shamir,  } {\it Lattice Chiral Fermions
  }{ Nucl.  Phys.  Proc.  Suppl.  {\bf } 47 (1996) 212,  hep-lat/9509023;
V.~Furman, Y.~Shamir, Nucl.Phys. B {\bf 439 } (1995), hep-lat/9405004.}}

 \lref\kaplan{ {D.B.~Kaplan, }  {\it A Method for Simulating Chiral
Fermions on the Lattice,} Phys. Lett. B {\bf 288} (1992) 342; {\it Chiral
Fermions on the Lattice,}  {  Nucl. Phys. B, Proc. Suppl.  {\bf 30} (1993)
597.}}

\lref\neubergerr{ {H.~Neuberger, } {\it Chirality on the Lattice},
hep-lat/9808036.}

\lref\neubergers{ {Rajamani Narayanan, Herbert Neuberger,} {\it INFINITELY MANY
    REGULATOR FIELDS FOR CHIRAL FERMIONS.}
    Phys.Lett.B302:62-69,1993.
    [HEP-LAT 9212019]}

\lref\neubergert{ {Rajamani Narayanan, Herbert Neuberger,}{\it CHIRAL FERMIONS
    ON THE LATTICE.}
    Phys.Rev.Lett.71:3251-3254,1993.
    [HEP-LAT 9308011]}

\lref\neubergeru{ {Rajamani Narayanan, Herbert Neuberger,}{\it A CONSTRUCTION
OF LATTICE CHIRAL GAUGE THEORIES.}
    Nucl.Phys.B443:305-385,1995.
    [HEP-TH 9411108]}

\lref\neubergerv{ {Herbert Neuberger,}{\it EXACTLY MASSLESS QUARKS ON THE
    LATTICE.}
    Phys. Lett. B417 (1998) 141-144.
    [HEP-LAT 9707022]}

%The first 3 papers deal with chiral fermions in general, while the last
%with the
%particular case of vector like fermions. All these papers are quite well
%known.
%
%If you wish to quote reviews, the review by Shamir is seriously flawed.
%More recent
%reviews are available.  Surprisingly, I happen to like:

\lref\neubergerw{ {Herbert Neuberger,}{\it CHIRAL FERMIONS ON THE
LATTICE.}
    Nucl. Phys. B, Proc. Suppl. 83-84 (2000) 67-76.
    [HEP-LAT 9909042]}

\lref\zbgr {L.~Baulieu and D. Zwanziger, {\it QCD$_4$ From a
Five-Dimensional Point of View},    Nucl. Phys. {\bf B581} 2000, 604.}

\lref\bgz {L.~Baulieu, P. A. Grassi and D. Zwanziger, {\it Gauge and
Topological Symmetries in the Bulk Quantization of Gauge Theories},
Nucl. Phys. {\bf B597} 583-614, 2001.}

\lref\bulkq {L.~Baulieu and D. Zwanziger, {\it From stochastic
quantization to bulk quantization; Schwinger-Dyson equations and
the S-matrix},    JHEP 0108:016, 2001.}

\lref\bulkqg {L.~Baulieu and D. Zwanziger, {\it Bulk quantization of gauge
theories: confined and Higgs phases},    JHEP 0108:015, 2001.}

\lref\equivstoch{L.~Baulieu and D. Zwanziger, {\it Equivalence of
stochastic quantization and the Faddeev-Popov Ansatz},
Nucl. Phys. B193 (1981) 163-172.}

 \lref\zbsd {L.~Baulieu and D. Zwanziger, {
\it From stochastic quantization to bulk quantization: Schwinger-Dyson
equations and S-matrix QCD$_4$}, hep-th/0012103.}

\lref\cuzwns {A.~Cucchieri and D.~Zwanziger, {\it Numerical study of
gluon propagator and confinement scenario in minimal Coulomb gauge},
hep-lat/0008026.}

\lref\fitgrib {A.~Cucchieri and D.~Zwanziger, {\it Fit to gluon
propagator and Gribov formula},    hep-th/0012024.}

\lref\vanish {D.~Zwanziger, {\it Vanishing of zero-momentum lattice
gluon propagator and color confinement},   Nucl. Phys. {\bf B364}
(1991) 127-161.}

\lref\gribov {V.~N.~Gribov, {\it Quantization of non-Abelian gauge
theories},   Nucl. Phys. {\bf B139} (1978)~1-19.}

\lref\singer {I.~Singer, {\it }   Comm. Math. Phys. {\bf 60}
(1978)~7.}

\lref\feynman {R.~P.~Feynman, {\it The qualitative behavior of
Yang-Mills theory in 2+1 dimensions},   Nucl. Phys. {\bf B188} (1981)
479-512.}

\lref\cutkosky {R.~E.~Cutkosky, {\it}   J. Math. Phys. {\bf 25} (1984)
939; R. E. Cutkosky and K. Wang, Phys. Rev. {\bf D37} (1988) 3024; R. E.
Cutkosky, Czech. J. Phys. {\bf 40} (1990) 252.}

\lref\vanbaal{J. Koller and P. van Baal, Nucl. Phys. {\bf B302} (1988)
1; P. van Baal, Acta Phys. Pol. {\bf B20} (1989) 295;
P. van Baal, Nucl. Phys. {\bf B369} (1992) 259;
P. van Baal and N. D. Hari Dass, Nucl. Phys. {\bf B385} (1992) 185.}

\lref\smekal{L.~von Smekal, A.~Hauck and R.~Alkofer,  {\it A Solution to
Coupled Dyson-Schwinger Equations in Gluons and Ghosts in Landau Gauge,}  
Ann. Phys. {\bf 267} (1998) 1; L. von Smekal, A. Hauck and R. Alkofer, {\it The
Infrared Behavior of Gluon and Ghost Propagators in Landau Gauge QCD,}  
Phys. Rev. Lett. {\bf 79} (1997) 3591; L. von Smekal {\it Perspectives for
hadronic physics from Dyson-Schwinger equations for the dynamics of quark and
glue,} Habilitationsschrift, Friedrich-Alexander Universit\"{a}t,
Erlangen-N\"{u}rnberg (1998).}

\lref\smekrev {R.~Alkofer and L.~von Smekal, {\it The infrared behavior of
QCD Green's functions},   Phys. Rep. {\bf 353} 281-465, 2001.}

\lref\lerche {C. Lerche, 
%{\it German title},   
Diplomarbeit, Friedrich-Alexander Universit\"{a}t, Erlangen-N\"{u}rnberg
(2001), http://theorie3.physik.uni-erlangen.de/~smekal/clerche.ps.gz.}

\lref\atkinsona {D.~Atkinson and J.~C.~R.~Bloch, {\it Running coupling in
non-perturbative QCD,}   Phys. Rev. {\bf
D58} (1998) 094036.}

\lref\atkinsonb {D.~Atkinson and J.~C.~R.~Bloch, {\it QCD in the infrared with
exact angular integrations,}   Mod. Phys. Lett. {\bf A13} (1998) 1055.}

\lref\brown {N.~Brown and M.~R. Pennington, {\it}   Phys. Rev. {\bf D38} (1988)
2266; Phys. Rev. {\bf D39} (1989) 2723.}

\lref\szczep {A.~P.~Szczepaniak and E.~S.~Swanson, {\it Coulomb Gauge QCD,
Confinement, and the Constituent Representation},   hep-ph/0107078.}

%%%%%%

%\draft

%%%%%%%%%

\Title{\vbox
{\baselineskip 10pt
\hbox{hep-th/0109224}
%\hbox{CERN-TH-00-??}
%\hbox{LPTHE-00-50}
\hbox{NYU-ThPh-9.28.01}
 \hbox{   }
}}
{\vbox{\vskip -30 true pt
\centerline{
   }
\medskip
 \centerline{  }
\centerline{Non-perturbative Landau gauge}
\centerline{and infrared critical exponents in QCD}
\medskip
\vskip4pt }}
\centerline{
%{\bf Laurent Baulieu}$^{  \dag     }$ and  
{\bf  Daniel Zwanziger}
%$^{ \ddag}$
}
\centerline{
%baulieu@lpthe.jussieu.fr, 
Daniel.Zwanziger@nyu.edu}
%pag5@nyu.edu
\vskip 0.5cm
%\centerline{\it $^{\dag}$LPTHE, Universit{\'e}s P. \& M. Curie (Paris~VI)
%et D. Diderot (Paris~VII), Paris,  France,}
%{\foot{UMR 7589 associ{\'e}e CNRS et
%Universit{\'e}s P. \&M. Curie (Paris~VI) et D. Diderot (Paris~VII)},
%Boite 126,
%4 place Jussieu, F-75252
%Paris Cedex 05, France.}
%\centerline{\it $^{\dag}$  Dept. of Physics, University
%of Rutgers, New Brunswick, NJ 60637, USA }
%\centerline{\it $^{\S}  $}
\centerline{\it 
%$^{\ddag}$   
Physics Department, New York University,
New-York,  NY 10003,  USA}

\medskip
\vskip  1cm
\noindent

	We discuss Faddeev-Popov quantization at the non-perturbative level and
show that Gribov's prescription of cutting off the functional integral at the
Gribov horizon does not change the Schwinger-Dyson equations, but rather
resolves an ambiguity in the solution of these equations.  We note that
Gribov's prescription is not exact, and we therefore turn to the method of
stochastic quantization in its time-independent formulation, and
recall the proof that it is correct at the non-perturbative level.  The
non-perturbative Landau gauge is derived as a limiting case, and it is found
that it yields the Faddeev-Popov method in Landau gauge with a cut-off at the
Gribov horizon, plus a novel term that corrects for over-counting of Gribov
copies inside the Gribov horizon.  Non-perturbative but truncated coupled
Schwinger-Dyson equations for the gluon and ghost propagators $D(k)$ and $G(k)$
in Landau gauge are solved asymptotically in the infrared region.  The infrared
critical exponents or anomalous dimensions, defined by $D(k) \sim 1/(k^2)^{1 +
a_D}$ and $G(k) \sim 1/(k^2)^{1 + a_G}$ are obtained in space-time dimensions
$d = 2, 3, 4$.  Two possible solutions are obtained with the values, in $d =
4$ dimensions,  
$a_G = 1, \ a_D = -2$, or 
$ a_G = (93 - \sqrt1201)/98 \approx 0.595353, \ a_D = - 2a_G$.

\Date{\ }

\def\e{\epsilon}
\def\demi{{1\over 2}}

\def\a{\alpha}
\def\b{\beta}
\def\d{\delta}

\def\m{\mu}
\def\n{\nu}

\def\l{\lambda}

\def\o{\omega}

\def\k{\kappa}

\newsec{Introduction}

	The problem of confinement in QCD presents a challenge to the
theorist.  One would like to understand how and why QCD describes a
world of color-neutral hadrons with a mass gap, even though it appears
perturbatively to be a theory of unconfined and massless gluons and
quarks.  A basic insight into the origin of the mass gap in
gluodynamics was provided by Feynman \feynman, Gribov \gribov, and
Cutkosky \cutkosky.  These authors proposed that the mass gap is
produced by the drastic reduction of the physical configuration space in
non-Abelian gauge theory that results from the physical identification,
$A_1 \sim A_2$, of distinct but gauge-equivalent field configurations,
$A_2 = {^g}A_1$.  A simple analogy is the change in the spectrum of
a free particle moving on a line, when points on the line are identified
modulo~$2\pi$.  The real line is reduced to the circle so the continuous
spectum becomes discrete.  In analytic calculations, the appropriate
identification of gauge-equivalent configurations requires
non-perturbative gauge-fixing, and in this article we approach the
confinement problem by considering the how non-perturbative
gauge-fixing impacts the Schwinger-Dyson equations of gluodynamics and
their solution.  We discuss both the Faddeev-Popov formulation, for
which the Gribov problem may have an approximate -- but not an exact --
solution, and stochastic gauge fixing which overcomes this difficulty
\dan.  We also briefly compare our results for the infrared critical
exponents with numerical evaluations of QCD propagators.  In this
connection we note that stochastic gauge fixing of the type considered
here has been implemented numerically with good statistics on
impressively large lattices  ($48^3 \times 64$)
\nakamuraa, \nakamurab, \nakamurac, and \nakamurad.  This opens
the exciting perspective of close comparison of analytic and numerical
calculations in this gauge.  Stochastic quantization has also been
adapted  to Abelian projection \nakamurae, and convergence of the
stochastic process has been studied theoretically~\direnzo.

	We briefly review Faddeev-Popov quantization as a non-perturbative
formulation.  We note that the Faddeev-Popov weight $P_{\rm FP}(A)$
possesses {\it nodal surfaces} in $A$-space where the Faddeev-Popov determinent
vanishes, and that a cut-off of the functional integral on a nodal surface does
not alter the Schwinger-Dyson equations, because it does not introduce boundary
terms.   As a result, Gribov's prescription to cut off the functional integral
at the (first) Gribov horizon \gribov, a nodal a surface that completely
surrounds the origin \dzgribreg, does not change the Schwinger-Dyson equations
of Faddeev-Popov theory at all, but rather it resolves an ambiguity in the
solution of these equations.

  We recall that Gribov's prescription is not in fact exact because there are
Gribov copies inside the Gribov horizon that get over-counted.  We then turn to
the method of stochastic quantization as described by a time-independent
diffusion equation in $A$-space (so there is no fictitious ``fifth time")
\dan.  This method by-passes the Gribov problem of choosing a representative on
each gauge orbit because gauge-fixing is replaced by the introduction of a
``drift force" that is the harmless generator of a gauge transformation.  We
next derive a formulation of the Landau gauge, that is valid
non-perturbatively, as a limiting case of stochastic quantization.  It yields
the Faddeev-Popov theory with a cut-off at the Gribov horizon, plus a novel
term the corrects for over-counting inside the Gribov horizon.  

	However attractive a formulation may be that is valid at the
non-perturbative level, it would remain largely ornamental without actual
non-perturbative calculations.  Fortunately, progress in finding approximate but
non-perturbative solutions for the propagators in QCD has been achieved
recently within the framework of Faddeev-Popov theory both in Coulomb gauge
using the hamiltonian formalism~\szczep, and in Landau gauge by solving a
truncated set of Schwinger-Dyson equations~\smekal, \atkinsona, and
\atkinsonb.  The Schwinger-Dyson approach is reviewed in~\smekrev.  In
the latter part of the present article we solve the Schwinger-Dyson equations
in the non-perturbative Landau gauge to obtain the infrared critical exponents
or anomalous dimensions of the gluon and ghost propagators $D(k)$ and $G(k)$ in
$d = 2, 3$ and 4 space-time dimensions.  The novel term is ignored here, in
order to compare with other recent calculations, but we explicitly select the
solution to the Schwinger-Dyson equations that vanishes outside the Gribov
horizon. Although a truncation is necessarily required to solve these equations,
nevertheless the values obtained for the infrared asymptotic dimensions agree
with exact results for probability distributions that vanish outside the Gribov
horizon \vanish, namely the vanishing of $D(k)$ at $k = 0$, and an enhanced
infrared singularity of $G(k)$. These properties also characterize the
non-perturbative solutions of the Schwinger-Dyson equations in QCD
obtained in recent studies, ~\smekal, \atkinsona, and \atkinsonb, and we verify
that they have also adopted the solution that vanishes outside the Gribov
horizon.

\newsec{Faddeev-Popov quantization at the non-perturbative level}

	The standard Faddeev-Popov Euclidean weight in Landau gauge is given by
\eqn\fadpop{\eqalign{
P_{\rm FP}(A) & = Q_{\rm FP}(A^{\rm tr}) \ \d(\p_\m A_\m)    \cr
Q_{\rm FP}(A^{\rm tr}) & = N\exp[-S_{\rm YM}(A^{\rm tr})] 
\det[-\p_\m D_\m(A^{\rm tr})] }}
with partition function 
\eqn\partfunc{\eqalign{
Z(J) & = \int dA \ P_{\rm FP}(A) \  \exp (J, A)    \cr
& = \int dA^{\rm tr} \ Q_{\rm FP}(A^{\rm tr}) \  \exp (J, A^{\rm tr}), 
}}
where $(J,A) \equiv \int d^4x J_\m^a A_\m^a$.  It depends only on the transverse
components of $J$ and we write $Z = Z(J^{\rm tr})$. 
The complete set of Schwinger-Dyson (SD) equations reads 
\eqn\sdfadpop{\eqalign{
{{\d \Sigma} \over {\d A^{\rm tr}}} (\d/\d J^{\rm tr}) \ Z = J^{\rm tr} \ Z, 
}}
where 
$\Sigma(A^{\rm tr}) \equiv 
S_{\rm YM}(A^{\rm tr}) - {\rm Tr} \ln[-\p_\m D_\m(A^{\rm tr})]$.

	The Faddeev-Popov operator 
$-\p_\m D_\m^{ac}(A^{\rm tr}) 
= - \p^2 \d^{ac} - f^{abc}A_\m^{{\rm tr},b} \p_\m$ is
hermitian because $A^{\rm tr}$ is transverse, $\p_\m A_\m^{\rm tr} = 0$. 
However it is not positive for every $A^{\rm tr}$. Because the Faddeev-Popov
determinent is the product of non-trivial eigenvalues, 
$\det[-\p_\m D_\m(A^{\rm tr})] = \prod_n \l_n(A^{\rm tr})$, it vanishes
together with $Q_{\rm FP}(A^{\rm tr})$ whenever any eigenvalue vanishes, and the
equation  $\l_n(A^{\rm tr}) = 0$ defines a {\it nodal surface} of 
$Q_{\rm FP}(A^{\rm tr})$ in $A^{\rm tr}$-space.  The nodal surface where the 
lowest non-trivial eigenvalue vanishes, defined by $\l_0(A^{\rm tr}) = 0$,
defines what is known as the (first) Gribov horizon.  It forms the boundary of
the region $\Omega$, known as the Gribov region, with the defining property that
all non-trivial eigenvalues of the Faddeev-Popov operator 
$-\p_\m D_\m(A^{\rm tr})$ are positive.  Clearly $Q_{\rm FP}(A^{\rm tr})$ is 
positive inside $\Omega$ and vanishes on  $\p \Omega$.  It is known that the
Gribov region has the following 3 properties:  (i) it is a convex region of
$A$-space, (ii) it is bounded in every direction, and (iii) it includes the
origin \dzgribreg. 

	The existence of nodal surfaces of $Q_{\rm FP}(A^{\rm tr})$ implies that the 
solution of the Schwinger-Dyson (SD) equation for $Z(J^{\rm tr})$ is not
unique.  For if the integral \partfunc\ that defines $Z(J^{\rm tr})$ is cut-off
on any nodal surface, the same SD equation follows, without any boundary
contribution.  Moreover since the SD equation for $Z(J^{\rm tr})$ is linear,
any linear combination of two solutions is a solution.  These ambiguities are
reflected in corresponding ambiguities in the solution of the SD equation for 
$W(J^{\rm tr}) \equiv \ln Z(J^{\rm tr})$, and for the effective action
$\Gamma(A^{\rm tr})$, obtained from $W(J^{\rm tr})$ by Legendre transformation.

	[We illustrate these points by a baby model.  Replace the field 
$A_\m^{\rm tr}(x)$ by a real variable $a$, and the Faddeev-Popov weight $Q_{\rm
FP}(A^{\rm tr})$ by 
$p(a) \equiv \exp(-\demi a^2) (1-g^2a^2)$, which vanishes on the ``Gribov
horizon" $a = \pm g^{-1}$.  The partition function 
$Z_1(j) 
\equiv \int_{-\infty}^{+\infty}da \ p(a) \exp(ja)$ satisfies the SD
equation 
\eqn\sdmod{\eqalign{
{ {\p \Sigma} \over {\p a} }\Big( { {\p } \over {\p j} }\Big) \ Z_1(j)= 
\Big[{ {\p } \over {\p j} } 
- g\Big(1+g{ {\p } \over {\p j} } \Big)^{-1} 
+ g\Big( 1 - g{ {\p } \over {\p j} } \Big)^{-1} \Big] \  Z_1(j) = j \ Z_1(j),
}}
corresponding to the action $\Sigma = \demi a^2 - \ln(1-g^2a^2)$. 
Suppose we restrict the integral to the Gribov region, $|a| \leq g^{-1}$,
so the partition function is given instead by
$Z_2(j) = \int_{-1/g}^{1/g} da \ p(a)\exp(ja)$.
The change occurs only for 
$a^2 > 1/g^2$, so $Z_1(j)$ and $Z_2(j)$ have the
same perturbative expansion.  It is clear that  $Z_1(j)$ and $Z_2(j)$ satisfy
the {\it same} SD equation \sdmod\ without a boundary contribution,
because $p(a)$ vanishes on the boundary $a = \pm 1/g$.  
Moreover, because the SD equation for $Z(j)$ is linear,
any linear combination, $Z(j) = \a  Z_1(j) + \b Z_2(j)$ also satisfies the
same SD equation.  Of course only one of them corresponds to a weight that
is positive everywhere.  This example easily generalizes to any number of
dimensions.]

	Gribov proposed to cut off the integral on the boundary $\p
\Omega$ of what we now call the Gribov region $\Omega$, so the partition
function is given by 
\eqn\cutpartf{\eqalign{ 
Z_\Omega(J^{\rm tr}) \equiv \int_\Omega dA^{\rm tr} 
\ Q_{\rm FP}(A^{\rm tr}) \exp(J^{\rm tr},A^{\rm tr}), 
}}  
for he conjectured that the region where the Faddeev-Popov operator is positive
contains only one gauge copy on each gauge orbit.  Since $\p \Omega$ is a nodal
surface, $Z_\Omega$ satisfies the {\it same} Schwinger-Dyson equation
\sdfadpop,
${{\d \Sigma} \over {\d A^{\rm tr}}} (\d/\d J^{\rm tr}) 
Z_\Omega = J^{\rm tr} Z_\Omega$.
Instead, Gribov's proposal selects a particular one out of a class
of non-perturbative solutions of these equations.\foot{For
partition function $Z_\Omega$, the expectation value of
$A^{\rm tr}$ in the presence of the source $J^{\rm tr}$ is 
$A_{\rm cl} \equiv \langle A^{\rm tr} \rangle_{J^{\rm tr}} 
 = \d W_\Omega/ \d J^{\rm tr}$, 
where $W_\Omega = \ln Z_\Omega$.  Because
the probability distribution in the presence of sources 
$Q(A^{\rm tr}) \exp(J^{\rm tr},A^{\rm tr})$ 
is positive in the Gribov region $\Omega$, and
because $\Omega$ is convex, it follows that $A_{\rm cl}$ lies in $\Omega$. 
Consequently the effective action
$\Gamma_\Omega(A_{\rm cl})$, obtained by Legendre transform from
$W_\Omega$, is defined only on the Gribov region.}

	It is now known however that Gribov's conjecture is not exact.  Indeed, there
are Gribov copies inside the Gribov horizon.\foot{As shown in sec.~4, they are
given by ${^gA}$, where $g = g_{\rm min}(x)$ is any local minimum of the
functional
$F_A(g) = \int d^4x \ |{^g A}|^2$, where ${^g A = g^{-1}Ag + g^{-1}\p g}$ 
is the transform of $A$ by the local gauge transformation $g(x)$.
In a lattice discretization, the link variable corresponding to the field
$A(x)$ is  generically a random field, so the minimization problem is of
spin-glass type which is known to have many solutions.  On the other hand for a
smooth configuration, such as the vacuum, $A = 0$, there are few solutions. 
Thus the number of copies is different for different orbits.  Moreover, since
the Faddeev-Popov weight is positive inside the Gribov horizon, there can be no
cancellations to save the day.  Note also that, in a lattice discretization,
the variables that characterize a configuration take values in a compact
space, so a minimizing configuration always exists, which shows that 
$\Omega$ contains at least one Gribov copy for each orbit.}  We shall
show however that an exact non-perturbative formulation yields Gribov's
proposal plus a well-defined correction term that corrects for overcounting
inside the Gribov horizon.

	There is an alternative proposal to make the Faddeev-Popov method valid
non-perturbatively.  It is conjectured that if one sums over all Gribov copies
using the signed Faddeev-Popov determinent $\det[-\p_\m D_\m(A^{\rm tr})]$, then
additional Gribov copies cancel in pairs, the reason being that the signed
determinent counts the signed intersection number which is a topological
invariant.  This is presumably the outcome of BRST quantization which has
formal properties that suggest it may be valid non-perturbatively \lbpert. 
Moreover this conjecture is supported by simple models \hirschfeld\ and
\fleepr.  However it is not known at present how to turn this prescription into
a non-perturbative calculational scheme in QCD, for example, by selecting a
particular solution of the SD equations, and moreover if the measure is not
everywhere positive, there is the danger of delicate cancellations that may
cause an approximate solution to be unreliable.  On the other hand, the
Gribov proposal is easily implemented, for example by requiring 
that the solution of the SD equation possess positivity properties.

\newsec{Time-independent stochastic quantization}

	The difficulties with Faddeev-Popov gauge fixing pointed out by Gribov are
by-passed by stochastic or bulk quantization. This method is a
formalization of a Monte Carlo simulation \pawu, and in its most
powerful formulation it makes use of a fictitious ``fifth
time'' that corresponds to computer time or number of sweeps of the lattice
in a Monte Carlo simulation.  Despite the Gribov ambiguity, there is no problem
of over-counting with gauge fixing in Monte Carlo simulations, which is achieved
by a gauge transformation of choice after any sweep, nor is there one in
stochastic or bulk quantization which relies on an infinitesimal gauge
transformation.  In the 5-dimensional
formulation, the field $A_\m^a = A_\m^a(x,t)$ depends on the 4 Euclidean
coordinates $x_\m$ and on the fifth time $t$.  One may easily write down the SD
equations for this formulation which involves a local 5-dimensional action, and
BRST invariances and Slavnov-Taylor identities are available to control
divergences \zbgr, \bgz, and \bulkqg.  However the 5-dimensional propagator 
$D = D(k^2, \o)$ depends on two invariants
$k^2$ and $\omega$, which makes the solution of the SD equations in 5
dimensions more complicated, and we shall not attempt it here.

	Instead we turn to an older 4-dimensional formulation of stochastic
quantization \dan.  It is based on an analogy between the (formal) Euclidean
weight $P_{\rm YM}(A) = \exp[- S_{\rm YM}(A)]$ and the Boltzmann distribution
$P(x) = \exp[-V(x)]$.  The latter  is the solution of the time-independent
diffusion equation
${{\p} \over {\p x_i}}({{\p} \over {\p x_i}} + {{\p V} \over {\p x_i}} ) P = 0$,
where the drift force is 
$K_i = - {{\p V} \over {\p x_i}}$.  We shall shortly consider more general
drift forces $K_i$ that are not necessarily conservative.  The field
theoretic analog of this equation is
\eqn\diffeq{\eqalign{
H_{\rm YM}P(A) \equiv - \int d^4x \ {{\d} \over {\d A_\m(x)}}
\Big( {{\d} \over {\d A_\m(x)}} + {{\d S_{\rm YM}} \over {\d A_\m(x)}} \Big)
P(A)  = 0, 
}}
where the drift force is
$K_{{\rm YM},\m}(x) \equiv - {{\d S_{\rm YM}} \over {\d A_\m(x)}}$, which is
solved by $P(A) = \exp(- S_{\rm YM})$.

	For a gauge theory, this solution is not normalizable.  However for a gauge
theory, one may modify the drift force 
$K_{\rm YM} \to K_{\rm YM} + K_{\rm gt}$
by adding to it a ``force" $K_{\rm gt}$ tangent to
the gauge orbit, without changing the expectation-value of gauge-invariant
observables $O(A)$. Such a force has the form of an infinitesimal gauge
transformation $K_{{\rm gt},\m}^a \equiv(D_\m v)^a$, where $v^a(x;A)$ is an
element of the Lie algebra, and
$(D_\m v)^a = \p_\m v^a + f^{abc}A_\m^b v^c$ is its gauge-covariant
derivative.  This force is not conservative, which means that it cannot be
expressed as a gradient, 
$K_{{\rm gt},\m} \neq - {{\d \Sigma} \over {\d A_\m}}$,
so this method is not available in a local action formalism in 4-dimensions. 
The total drift force is given by
\eqn\drift{\eqalign{
K_\m & \equiv K_{{\rm YM},\m} + K_{{\rm gt},\m}   \cr
	& = - {{\d S_{\rm YM}} \over {\d A_\m}} + D_\m v,
}}
and $P(A)$ is the solution of the
modified time-independent diffusion equation 
\eqn\diffeq{\eqalign{
HP & = (H_{\rm YM} + H_{\rm gt}) P    \cr
   & = - \int d^4x \ {{\d} \over {\d A_\m}}
\Big( {{\d} \over {\d A_\m}} + {{\d S_{\rm YM}} \over {\d A_\m}}
 - D_\m v \Big) \ P = 0.
}}
It is easy to show \dan\ that the expectation-value 
$\langle O \rangle = \int dA \ O(A) P(A)$, 
of gauge-invariant observables $O(A)$
is independent of $v$, using the fact
that $H_{\rm gt}^{\dag}$ is the generator of an infinitesimal local gauge
transformation 
\eqn\ggetrans{\eqalign{
H_{\rm gt}^{\dag} & = G(v) = \int d^4x \ v(x) G(x)   \cr
G(x) & \equiv D_\m {{\d} \over {\d A_\m}}
}}
and that $O(A)$ and $H_{\rm YM}$ are gauge invariant,
$G(x)O = 0$, and $[G(x), H_{\rm YM}] = 0$.

	The additional drift force $K_{{\rm gt}, \m} = D_\m v$ must be chosen so that
it is globally a restoring force along gauge orbits, thus preventing the
escape of probability to infinity along the gauge orbit where 
$S_{YM}$ is flat.  This may be achieved by choosing
$K_{\rm gt}$ to be in the direction of steepest descent, restricted to gauge
orbit dirctions, of some conveniently chosen minimizing functional $F(A)$.  A
convenient choice is the Hilbert square norm, 
$F(A) = ||A||^2 = \int d^4x |A_\m|^2$. For a generic infinitesimal
variation restricted to gauge orbit directions 
$\d A_\m = \e D_\m \o$, we have
\eqn\minfunc{\eqalign{
\d F(A) = 2 (A_\m, \d A_\m) = 2 (A_\m, \e D_\m \o) = 2\e (A_\m, \p_\m \o)
= - 2\e (\p_\m A_\m, \o ).
}}
The direction of steepest descent of $||A||^2$, restricted to gauge orbit
directions, is seen to be $\d A_\m = \e D_\m \o$, for $\o = \p_\l A_\l$. Thus if
we choose $v = a^{-1} \p_\l A_\l$, where $a$ is a positive gauge parameter, the
drift force $K_{{\rm gt},\m} \equiv a^{-1} D_\m \p_\l A_\l$ points globally in
the direction of steepest descent, restricted to gauge orbit directions, of the
minimizing functional $||A||^2$.  In the following we shall use the
time-independent diffusion equation
\eqn\diffeqa{\eqalign{
HP \equiv - \int d^4x \ {{\d} \over {\d A_\m}}
\Big( {{\d} \over {\d A_\m}} + {{\d S_{\rm YM}} \over {\d A_\m}}
 - a^{-1} D_\m \p_\l A_\l \Big) \ P = 0.
}}
The 5-dimensional formulation is based on the corresponding time-dependent
diffusion equation
\eqn\tddiffeq{\eqalign{
 \p P / \p t= - HP.
}}

\newsec{Non-perturbative Landau gauge}

	Because the gauge-fixing force points in the direction of steepest descent
of the minimizing functional $F(A) = ||A||^2$, restricted to gauge orbit
directions, it follows that for large values of the gauge parameter $a^{-1}$ the
probability gets concentrated near the local minima of this functional
restricted to gauge orbit variations.\foot{These conditions define a local
minimum at $g(x) = 1$ of the functional on the gauge orbit through $A$ defined
by $F_A(g) = \int d^4x |{^g}A|^2$.}  At a local minimum the first variation
vanishes for all $\o$, $\d F(A) = - 2\e (\o, \p_\m A_\m ) = 0$, 
as we have just seen, so at a
minimum, the Landau gauge condition $\p_\m A_\m = 0$, is satisfied.  {\it In
addition}, the second variation in gauge orbit directions is non-negative, 
$\d^2 F(A) = - 2\e (\o, \p_\m \d A_\m ) = - 2\e^2 (\o, \p_\m D_\m \o )
\geq 0$, for all $\o$, which is the statement that the Faddeev-Popov operator
$M(A) \equiv - \p_\m D_\m(A)$ is positive.  These are the
defining properties of the Gribov region, and we conclude that in the limit in
which the gauge parameter approaches zero, $a \to 0$, the probability $P(A)$
gets concentrated on transverse configurations $A = A^{\rm tr}$ that lie
{\it inside} the Gribov horizon.  We have noted above that there are Gribov
copies inside the Gribov horizon.  However the present method does not
require that the probability gets concentrated on any particular one of them
such as, for example, the absolute minimum of the minimizing function,
and for finite gauge parameter $a$, the gauge-fixing is ``soft" in the sense
that no particular gauge condition is imposed.  For gauge-invariant
observables, it does not matter how the probability is distributed
along a gauge orbit, but only that it be correctly
distributed between gauge orbits.  This is assured because a harmless gauge
transformation was introduced instead of gauge fixing.

	We have noted that $A$ becomes purely transverse in the limit $a \to 0$. 
We shall solve \diffeqa\ in this limit by the Born-Oppenheimer method in order
to obtain the non-perturbative Landau gauge.  For small $a$, the
longitudinal component of $A$ is small and, as we shall see, it evolves rapidly
compared to the transverse component.  However because of the factor $a^{-1}$ in
\diffeqa, the mean value of the longitudinal part of the gluon propagator
strongly influences the transverse propagator in the limit $a \to 0$. 

	We decompose $A$ into its transverse and longitudinal
parts according to
$A_\m^b = A_\m^{{\rm tr},b} + a^{1/2}\p_\m (\p^2)^{-1}L^b$, so
$\p_\m A_\m^b = a^{1/2} L^b$, and
${{\d} \over {\d A_\m^b}} = {{\d} \over {\d A_\m^{ {\rm tr},b} }} 
- a^{-1/2}\p_\m {{\d} \over {\d L^b}}$.  
In terms of these variables, eq. \diffeqa\ reads
\eqn\diffeqb{\eqalign{
 \int d^4x \ \Big[ &{{\d} \over {\d A_\m^{\rm tr} }}
\Big( {{\d} \over {\d A_\m^{\rm tr} }} 
+  {{ \d S_{\rm YM} } \over {\d A_\m}} 
 - a^{-1/2} (A_\m \times L) \Big)  \cr
&+ a^{-1}{{\d} \over {\d L}}
\Big( (-\p^2) {{\d} \over {\d L}} 
+ a^{1/2}\p_\m  {{\d S_{\rm YM}} \over {\d A_\m}} 
 - \p_\m D_\m(A) L \Big) \ \Big] \ P = 0,
}}
where we have used the notation $(K \times L)^b = [K,L]^b = f^{bcd}K^cL^d$,
for elements $K$ and $L$ of the Lie algebra.  The leading terms in $H$ are of
order $a^{-1}$, $a^{-1/2}$ and $a^0$, and we expand 
$P = P_0 + a^{1/2}P_1 + ... \ $.  The leading term, of order $a^{-1}$, reads
\eqn\diffeqlo{\eqalign{
 \int d^4x \  {{\d} \over {\d L}}
\Big( (-\p^2) {{\d} \over {\d L}} 
 - \p_\m D_\m(A^{\rm tr}) L \Big) \ P_0 = 0.
}}
This is solved by $P_0$ that is Gaussian in $L$, 
\eqn\sol{\eqalign{
P_0(A^{\rm tr}, L) = Q(A^{\rm tr}) \ (\det X)^{-1/2} \ \exp[- 1/2 (L, XL)],
}}
where $X^{bc}(x,y; A^{\rm tr})$ is a symmetric kernel.  Equation \diffeqlo\ is
satisfied provided $X$ satisfies
$(L, X(-\p^2)XL) - (L, XML) = 0$ identically for all $L$, 
and ${\rm tr}[(-\p^2)X - M] = 0$.  Here 
$M = M(A^{\rm tr}) \equiv - \p_\m D_\m(A^{\rm tr})$ is the Faddeev-Popov
operator that is symmetric for $A$ transverse.  The first equation yields
$X(-2\p^2)X = XM + MX$, or 
$MY + YM = -2\p^2$ for $Y \equiv X^{-1}$.  
Moreover when this equation is satisfied, it implies that the second equation is
also satisfied. The equation for $Y$ is linear.  To solve it we take
matrix elements in the basis provided by the eigenfunctions of the
Faddeev-Popov operator $M(A^{\rm tr}) u_n = \l_n u_n$,   where $\l_n =
\l_n(A^{\rm tr})$, and obtain
\eqn\sola{\eqalign{
(u_m, X^{-1}u_n) = (u_n, Yu_m) = (\l_m + \l_n)^{-1} (u_n, (-2\p^2)u_m).  
}} 
We see that the Gaussian solution $P_0(A^{\rm tr}, L)$ is normalizable in $L$
only when all the eigenvalues $\l_n(A^{\rm tr})$ are positive, namely, 
for $A^{\rm tr}$ inside the Gribov region.  However we have seen above that in
the limit $a \to 0$ the solution $P(A)$ is supported inside the Gribov
region.  Thus the coefficient function $Q(A^{\rm tr})$ carries a factor
$\theta(\l_0(A^{\rm tr}))$, that restricts the support of $P_0$ to this region.
Finally we note that for $A^{\rm tr}$ in the Gribov region, $Y$ may be
written
\eqn\solb{\eqalign{
X^{-1} = Y = \int_0^{\infty} dt \ \exp(-Mt)(-2\p^2) \exp(-Mt). 
}}
This representation shows explicitly that $X$ is a positive operator for
$A^{\rm tr}$ inside the Gribov region.

	To determine $Q(A^{\rm tr})$ we substitute \sol\ into \diffeqb, and integrate
over $L$.  This kills the term in ${{\d} \over {\d L}}$.  It also kills the
term of order $a^{-1/2}$ in $a^{-1/2}A_\m \times L 
= a^{-1/2}[A_\m^{\rm tr} + a^{1/2}\p_\m (\p^2)^{-1}L] \times L$
because this term is odd in $L$. This gives in
the limit $a \to 0$, the finite equation for $Q(A^{\rm tr})$,
\eqn\diffeqtr{\eqalign{
 \int d^4x \  {{\d} \over {\d A_\m^{\rm tr} }}
\Big( {{\d} \over {\d A_\m^{\rm tr} }} 
+  {{ \d S_{\rm YM} } \over {\d A_\m}}(A^{\rm tr}) 
 -  K_{{\rm gteff},\m}(A^{\rm tr}) \Big)   \ Q(A^{\rm tr}) = 0.
}}
Here $K_{\rm gteff}$
is the average over $L$ of the gauge-transformation force, with weight
$(\det X)^{-1/2} \ \exp[- 1/2 (L, XL)]$, namely
\eqn\avdrift{\eqalign{ 
K_{{\rm gteff},\m}^b(x; A^{\rm tr}) & \equiv
\langle f^{bcd}[\p_\m(\p^2)^{-1} L^c](x) L^d(x) \rangle  \cr
& = f^{bcd}\p_\m(\p^2)^{-1}Y^{cd}(x,y; A^{\rm tr})|_{y=x}  \cr
& =  \int_0^{\infty} dt \ 
f^{bcd}\p_\m(\p^2)^{-1}[\exp(-Mt)(-2\p^2) \exp(-Mt)]^{cd}(x,y)|_{y=x},
}} 
and we have used
$Y^{cd}(x,y; A^{\rm tr}) = \langle L^c(x) L^d(y) \rangle$.  
We now take the limit $a \to 0$ namely 
$P(A) \to \lim_{a \to 0}P(A) = \lim_{a \to 0}P_0(A)$.  
With $L = a^{-1/2}\p_\m A_\m$ this gives
\eqn\landprob{\eqalign{ 
	P(A) = Q(A^{\rm tr}) \d(\p_\m A_\m),
}}
where $Q(A^{\rm tr})$ is the solution of \diffeqtr.  This defines the
non-perturbative Landau gauge.  

	To exhibit the relation between the non-perturbative
Landau gauge and Faddeev-Popov theory, we decompose $K_{\rm gteff}$, given in
eq. \avdrift, according to
\eqn\avdrifta{\eqalign{ 
K_{\rm gteff} & =  K_1 + K_2   \cr
K_{1,\m}^b(x) 
& = - f^{bcd}\p_\m (M^{-1})^{cd}(x,y)|_{y=x} \cr 
K_{2,\m}^b(x) & = 
\int_0^{\infty} dt \ 
f^{bcd}\p_\m(\p^2)^{-1}\{[2\p^2,\exp(-Mt)] \exp(-Mt)\}^{cd}(x,y)|_{y=x}.
}} 
The first term may be written
\eqn\avdrifto{\eqalign{ 
K_{1,\m}^b(x)  
= { { \d ({\rm tr}\ln M) } \over {\d A_\m^{ {\rm tr},b}(x) } } 
= - { { \d \Sigma } \over {\d A_\m^{ {\rm tr},b}(x) } } ,
}}
so $K_1$ is a conservative drift force, derived from an
action $\Sigma \equiv {\rm tr}\ln M = - (\ln \det M)$ that precisely reproduces
the Faddeev-Popov determinant.  So if $K_2$ were neglected we regain
the Faddeev-Popov theory, with the added stipulation to choose the solution that
vanishes outside the Gribov horizon.

	The second term may be simplified using the identity
\eqn\ident{\eqalign{ 
[ \p^2, \exp(-Mt)] = - \int_0^t ds \ \exp(-Ms)[ \p^2, M] \exp[-M(t-s)],
}}
which gives
\eqn\avdriftt{\eqalign{ 
K_{2,\m}^b(x)  = 
- \int_0^{\infty} ds \ 
f^{bcd}  \p_\m(\p^2)^{-1}\{ \exp(-Ms)[\p^2, M]
\exp(-Ms)M^{-1}\}^{cd}(x,y)|_{y=x}, }} 
where, $M = M(A^{\rm tr})$.  The ``drift force" $K_2$ is a novel term.  It's
presence is required to correct the overcounting, discussed in sec.~2,
that occurs when the Faddeev-Popov theory is cut off at the Gribov
horizon.\foot{In our derivation we used the Born-Oppenheimer method that
is non-perturbative in $g$ in order to obtain the $a \to 0$ limit at finite
$g$.  So the presence of the new term $K_2$ is not in contradiction with the
fact that the Faddeev-Popov theory provides a formal perturbative expansion that
has all the correct properties including perturbative unitarity.}

\newsec{Schwinger-Dyson equations}

	The partition function is defined by
\eqn\partfunland{\eqalign{
Z(J) & = \int dA \ Q(A^{\rm tr}) \d(\p_\m A_\m) \exp(J,A)  \cr
		    & = \int dA^{\rm tr} \ Q(A^{\rm tr}) \exp(J,A^{\rm tr}).}}
It depends only on the transverse component $J_\m^{\rm tr}$ of $J_\m$
(on-shell gauge condition), and we write $Z = Z(J^{\rm tr})$.
Generally, in the Faddeev-Popov approach,
one relaxes the transversality condition, by writing
 $\d(\p_\m A_\m) = \int db \exp(i\int d^4x \ b \ \p_\m A_\m)$,
and then one uses Slavnov-Taylor identities to determine longitudinal parts
of vertices.  However these identities have not yet been derived in the
present 4-dimensional stochastic approach, and we shall solve the SD equations
using the on-shell formalism for the gauge-condition.  The on-shell
correlation functions, such as propagators, are the same as the off-shell ones,
but the vertices (one-particle irreducible functions) are strictly transverse. 
Renormalization theory is not well articulated at present in the on-shell
formalism, but we shall not encounter ultraviolet divergences in the SD
equations in the infrared limit. Moreover, we shall see that in this limit the
SD equations are invariant under the renormalization-group.\foot{It should also
be noted that it is not known at present how to maintain the
Slavnov-Taylor identities exactly at the non-perturbative level in the
off-shell formalism, although methods for dealing with this have been
proposed \brown.}

	The partition function $Z(J^{\rm tr})$, which is the generating
functional of (transverse) correlation functions, is the fourier transform of
the probability distribution $Q(A^{\rm tr})$.  Consequently the SD equation for 
$Z(J^{\rm tr})$ is simply the diffusion equation \diffeqtr, expressed in terms
of the fourier-transformed variables,
\eqn\sdz{\eqalign{
 \int d^4x \  J^{\rm tr}_\m
\Big[ J^{\rm tr}_\m
 +  K_{{\rm toteff},\m}\Big({{ \d } \over {\d J^{\rm tr}}}\Big) \Big]   
\ Z(J^{\rm tr}) = 0.
 }}
Here we have introduced the total effective drift force
\eqn\ktotal{\eqalign{
K_{{\rm toteff},\m}(A^{\rm tr}) \equiv
 - {{ \d S_{\rm YM} } \over {\d A_\m}}(A^{\rm tr}) 
 +  K_{{\rm gteff},\m}(A^{\rm tr}).
}}
Only the transverse component of $K_{{\rm toteff},\m}$ appears in the following.
The free energy $W(J^{\rm tr}) \equiv \ln Z(J^{\rm tr})$, 
which is the generating functional of connected correlation functions,
satisfies the SD
equation
\eqn\sdw{\eqalign{
 \int d^4x \  J^{\rm tr}_\m
\Big[ J^{\rm tr}_\m
 +  K_{{\rm toteff},\m}
\Big({{ \d W(J^{\rm tr}) } \over {\d J^{\rm tr}}} 
+ {{ \d } \over {\d J^{\rm tr}}}\Big) \Big]   = 0.
 }}
The effective action is obtained by Legendre transformation,
$\Gamma(A^{\rm tr}) = (J^{\rm tr}, A^{\rm tr}) - W(J^{\rm tr})$ by inverting 
$A^{\rm tr} = {{ \d W } \over {\d J^{\rm tr}_\m}}$. 
It satisfies the SD equation
\eqn\sdg{\eqalign{
 \int d^4x \  {{ \d \Gamma(A^{\rm tr}) } \over {\d A_\m^{\rm tr} }}
\Big[ {{ \d \Gamma(A^{\rm tr}) } \over {\d A_\m^{\rm tr}}}
 +  K_{{\rm toteff},\m}
\Big( A^{\rm tr} 
+ {\cal D}(A^{\rm tr}){{ \d } \over {\d A^{\rm tr}}}\Big) \Big]  = 0,
 }}
where the argument of $K_{\rm toteff}$ is written in matrix notation, and is given
explicitly by
$A_\m^{\rm tr}(x) + \int d^4y \ {\cal D}_{\m \n}(x,y;A^{\rm tr})
{{ \d } \over {\d A_\n^{\rm tr}(y)}}$.  Here
${\cal D}_{\m \n}(x,y;A^{\rm tr}) = 
{{ \d A_\n^{\rm tr}(y) } \over {\d J_\n^{\rm tr}(x)}}$ is the gluon propagator
in the presence of sources.  

	To obtain the SD equation for the propagator, we expand in powers of 
$A^{\rm tr}$,
\eqn\exppow{\eqalign{
{{ \d \Gamma(A^{\rm tr}) } \over {\d A_\m^{\rm tr} }} & = 
(D^{-1}A^{\rm tr})_\m + ...    \cr
K_{{\rm toteff},\m}^{\rm tr}
\Big( A^{\rm tr} + {\cal D}(A^{\rm tr}){{ \d } \over {\d A^{\rm tr}}}\Big) 
& = - (R A^{\rm tr})_\m + ... \ \ , 
}}
where we have again used matrix notation.  Here 
$D_{\m \n} = D_{\m \n}(x-y)$ is the gluon propagator in
the absence of sources, and
\eqn\defq{\eqalign{
R_{\m\n}(x-y) \equiv - \Big[{{ \d } \over {\d A_\n^{\rm tr}(y)}}
K_{{\rm toteff},\m}^{\rm tr}
\Big( A^{\rm tr} + {\cal D}(A^{\rm tr}){{ \d } \over {\d A^{\rm tr}}}\Big)(x)
\Big]|_{A^{\rm tr} = 0}. 
}}
Both $D$ and $R$ are identically transverse, and in momentum space,
by virtue of Lorentz invariance, are of the form 
$D_{\m\n}(k) = D(k^2) [\d_{\m\n} - k_\m k_\n/k^2]$ and
$R_{\m\n}(k) = R(k^2) [\d_{\m\n} - k_\m k_\n/k^2]$. 
Upon equating terms
quadratic in $A^{\rm tr}$ in \sdg\ we obtain
\eqn\quad{\eqalign{
\int d^4x \ [D^{-1}A_\m^{\rm tr}](x) \ [(D^{-1} - R)A_\m^{\rm tr}](x) = 0,
}}
which we write in matrix notation as
\eqn\matnot{\eqalign{
(D^{-1}A^{\rm tr}, [D^{-1} - R]A^{\rm tr}) = 0.
}}
This holds identically in $A^{\rm tr}$. From the expressions for 
$D_{\m \n}(k)$ and $R_{\m \n}(k)$, we
see that both are symmetric operators that commute, $DR = R D$.  As a result,
the operator appearing in the last equation is symmetric and must vanish,
\eqn\van{\eqalign{
D^{-1} (D^{-1} - R) = 0.
}}
This gives the SD equation for the gluon propagator
\eqn\sdprop{\eqalign{
D^{-1} = R,
}}
where $R$ is given in \defq.

\newsec{Solution of SD equation in the infrared}

	Recall the decomposition 
$K_{\rm gteff} = K_1 + K_2$, where 
$K_1  = { { \d ({\rm tr}\ln M) } \over {\d A^{\rm tr}(x) } }$
is the drift force that, in the absence of $K_2$, describes the
Faddeev-Popov theory in Landau gauge.  Since it is not without interest to
solve Faddeev-Popov theory non-perturbatively in Landau gauge,
and in order to compare our results with other authors,
we shall here ignore $K_2$, the novel term that corrects the over-counting that
occurs when the Faddeev-Popov theory is cut off at the Gribov horizon.  

	The remaining drift force, $K_1$, describes Faddeev-Popov theory in Landau
gauge.  We have seen in sec.~2 that there is an ambiguity in the solution of
the SD equations of the Faddeev-Popov theory, with no clear prescription to
resolve it at the non-perturbative level.  Fortunately the present derivation
provides the additional information that is needed to resolve this ambiguity:
{\it we must choose the solution of the SD equations that vanishes outside the
Gribov horizon} because, as we have seen, $Q(A^{\rm tr})$ vanishes outside the
Gribov horizon in the limit $a \to 0$.  With this choice it is likely that
qualitative features of the exact theory (with $K_2$) will be preserved.

 With neglect of $K_2$ we may write directly the
familiar SD equations of the Faddeev-Popov theory in Landau gauge, in an
arbitrary number $d$ of Euclidean dimensions
\eqn\sdgluon{\eqalign{
D_{\m \n}^{-1}(k) = \ & (\d_{\m \n}k^2 - k_\m k_\n)   \cr
& + N g^2 (2 \pi)^{-d} \int d^d p \ \ G(p+k) (p+k)_\m 
  \ G(p) \ \Gamma_\n(p,k)  + {\rm (gluon \  loops)} 
}}
\eqn\sdghost{\eqalign{
G^{-1}(p) =  p^2   
 - N g^2 (2 \pi)^{-d} \int d^d k 
 \ G(k + p) \ (k + p)_\m D_{\m \n}(k) \ \Gamma_\n(p,k),
}}
where $G(p)$ is the ghost propagator, $\Gamma_\n(p,k)$ is the full
ghost-ghost-gluon vertex.  In Landau
gauge, a factorization of the external ghost momentum occurs, so the
ghost-ghost-gluon vertex is of the form 
$\Gamma_\n(p,k) = \Gamma_{\n, \l}(p,k)p_\l$.  As a result there is no
independent renormalization of $\Gamma_\m(p,k)$, and the renormalization
constants in Landau gauge are related by 
$Z_g^2 Z_3 \tilde{Z}_3^2 = 1$, where $g_0 = Z_g g_r$, $D_0 = Z_3 D_r$, and
$G_0 = \tilde{Z}_3 G_r$.  So far, we have written the SD equations for
unrenormalized quantities, with the index $0$ suppressed.

	We must select the solution to these equations that corresponds to a
probability distribution $Q(A^{\rm tr})$ that vanishes outside the Gribov
horizon.  To do so, it is sufficient to impose any property that holds for
this distribution, provided only that it determines a unique
solution of the SD equations.  Besides positivity, which will be discussed in
the concluding section, there are two exact properties that hold for a
probability distribution $P(A^{\rm tr})$ that vanishes outside the Gribov
horizon: (i) the horizon condition, and (ii) the vanishing of the gluon
propagator at $k = 0$, \vanish\foot{The vanishing
of the gluon propagator at $k = 0$ results from the proximity of the Gribov
horizon in infrared directions.} 
\eqn\vanisheq{\eqalign{
\lim_{k \to 0} D(k) = 0.
}}
The horizon condition (i) is equivalent to 
the statement that $G(p)$ diverges more rapidly than~$1/p^2$, or
\eqn\horizony{\eqalign{
\lim_{p \to 0}[p^2 G(p)]^{-1} = 0.
}} 
Indeed if we divide the SD equation \sdghost\ by $p^2$, and impose this
conditon, we obtain  
\eqn\horizon{\eqalign{
\d_{\m \l} = N g^2 (2 \pi)^{-d} \int d^d k 
 \ G(k) \ D_{\m \n}(k) \ \Gamma_{\n \l}(0,k).
}}
This is the non-perturbative statement that the ghost self energy, which is of
the form 
$\Sigma(p) = p_\m \Sigma_{\m \l}(p) p_\l$ because of the
factorization of the external ghost momentum, exactly cancels the tree level
term at $p = 0$,
\eqn\horizonz{\eqalign{
\d_{\m \l} = \Sigma_{\m \l}(0).
}}
Equations \horizon\ and \horizonz\ are the form of the horizon condition
given in \horizcon, \horizpt, and \czcoulf.\foot{In a space of high dimension
$N$ the probability distribution within a smooth surface such as a sphere $r <
R$ gets concentrated near the surface $r = R$ because of the entropy or
phase-space factor $r^{N-1}dr$.  The horizon condition is the statement that
the probability distribution within the Gribov horizon is concentrated on the
Gribov horizon because the dimension $N$ of $A$-space diverges with the volume
$V$ in, say, a lattice discretization.} We will see that it is sufficient to
apply either condition (i) or (ii), and the other condition then follows
automatically.  The horizon condition allows us to write the SD equation for
the ghost propagator, $\sdghost$, in the form   
\eqn\sdghosta{\eqalign{
G^{-1}(p) =   
  N g^2 (2 \pi)^{-d} \int d^d k     
  \ p_\m D_{\m \n}(k) 
\ [\Gamma_{\n, \l}(0,k) \ G(k) - \Gamma_{\n, \l}(p,k) \ G(k + p)] p_\l ,
}}
where we have used $k_\m D_{\m \n}(k) = 0$.  This equation was solved
numerically in 3-dimenions in \czcoulf, using an assumed form for $D(k)$.

	We wish to determine the asymptotic form of the propators at low momentum,
$G^{\rm as}(p^2)$, and 
$D^{\rm as}_{\m \n}(k) = D^{\rm as}(k^2)P_{\m \n}^{\rm tr}(k)$, where
$P_{\m \n}^{\rm tr}(k) = \d_{\m \n} - k_\m k_\n/k^2$ is the transverse
projector.  For this purpose we let the external
momenta in the SD equations be asymptotically small compared to QCD mass
scales.  In this case the loop integration will be dominated by asymptotically
small loop momenta, so the propagators inside the integrals may also be
replaced by their asymptotic values.  This is true provided that the resulting
integrals converge, as will be verified.  We shall also truncate the SD
equations by neglecting transverse vertex corrections, as usual, in order to
obtain a closed system of equations, 
$\Gamma_\n^{\rm tr}(p,k) \to P_{\n \m}^{\rm tr}(k)p_\m$.  Such truncations may,
possibly, be justified {\it a posteriori} by calculating corrections to see if
they are small. Because $D_{\m \n}(k)$ is transverse, the SD equation for the
ghost propagator simplifies to
\eqn\sdghostb{\eqalign{
(G^{\rm as})^{-1}(p^2) =   
  N g^2 (2 \pi)^{-d} \int d^d k \ (k^2)^{-1} \ & [p^2 k^2 - (p\cdot k)^2] 
   D^{\rm as}(k^2)     \cr
& \times [\ G^{\rm as}(k^2) - \ G^{\rm as}((k + p)^2)].
}}
This equation is invariant under renormalization because of the identity
$Z_g^2 Z_3 \tilde{Z}_3^2 = 1$.   This allows us to take all quantities in the
last equation to be renormalized ones, with suppression of the index $r$.

	Because the asymptotic infrared limit is a critical limit, the asymptotic
propagators obey simple power laws,
\eqn\powerlaw{\eqalign{
D^{\rm as}(k^2) = c_D \m^{2\a_D}\ (p^2)^{-(1 + \a_D)}; \ \ \ \ \ \ \ \ \ \ \ 
g G^{\rm as}(p^2) = c_G \m^{2\a_G + (4-d)/2}\ (p^2)^{-(1 + \a_G)},
}}
according to standard renormalization-group arguments.  Here $\a_D$ and $\a_G$
are infrared critical exponents or anomalous dimensions that we shall
determine, while $\m$ is a mass scale, and $c_D$ and $c_G$ are dimensionless
parameters.  The horizon condition \horizony\ implies 
$\a_G > 0$, whereas \vanisheq\ implies $ \a_D <  -1$.  Upon changing
integration variable according to $k_\m = |p|k'_\m$, and equating like powers
of $p$ we obtain 
\eqn\powerlaw{\eqalign{
\a_D + 2\a_G =  - \ (4 - d)/2.
}}
The integral is ultraviolet convergent for 
$d - 2(1+\a_D) - 2(1+\a_G) - 2 < 0$, where 2 powers of $k$ are gained because
of the difference $[G^{\rm as}(k^2) - G^{\rm as}((k + p)^2)]$.  With  
$\a_D = - 2\a_G -  (4 - d)/2$, this gives  $\a_G < 1$ as the condition for
ultraviolet convergence, so $0 < \a_G < 1$.
  
	We now turn to the SD equation for the gluon propagator \sdgluon.  
In the exact Faddeev-Popov theory with off-shell gauge condition,
the right hand side of $\sdgluon$ is exactly transverse in $k$ on both free
Lorentz indices $\m$ and $\n$ by virtue of the Slavnov-Taylor identities.  This
allows us to apply transverse projectors 
$P_{\m' \m}^{\rm tr}(k) = \d_{\m\n} - k_\m k_\n/k^2$ and
$P_{\n' \n}^{\rm tr}(k)$ to these indices.  In our
derivation, with on-shell gauge condition, the projectors are automatically
applied.  As a result, since the gluon propagators are transverse, only the
transverse parts of the vertices contribute on the right-hand side.  We
therefore make the truncation approximation of replacing these transverse
vertices by their tree-level expressions.  We now estimate the various terms on
the right hand side of the SD equation \sdgluon\ for $D(k)$.  We just concluded
from the horizon condition and the SD equation for $G(p)$ that 
$\a_G > 0$ and $\a_D < 0$.  As a result, on the right-hand side of \sdgluon, the
ghost loop that we have written explicitly is more singular in the infrared
than the gluon loops.  Morover in the infrared, 
$D^{-1}(k) \sim (k^2)^{(1 + \a_D)}$ is more singular at $k = 0$ than the
tree-level term $\sim k^2$ because $\a_D < 0$.  We now let the external
momenutm $k$ have an asymptotically small value, so the loop integration is
dominated by asymptotically small values of the integration variable $p$
(provided the resulting integral converges).  We take the asymptotic infrared
limit of \sdgluon\  with external projectors and obtain
\eqn\sdgluona{\eqalign{
[D^{\rm as}(k^2)]^{-1}P_{\m \n}^{\rm tr}(k) =  N g^2 (2 \pi)^{-d} 
P_{\m \l}^{\rm tr}(k) 
\int d^d p \  p_\l  \ G^{\rm as}((k + p)^2) 
\ G^{\rm as}(p^2) \ p_\k \ P_{\k \n}^{\rm tr}(k). 
}} 
We take the trace on Lorentz indices and obtain
\eqn\sdgluonb{\eqalign{
[D^{\rm as}(k^2)]^{-1} =  N g^2 (2 \pi)^{-d} [(d-1)k^2]^{-1}
\int d^d p \  [p^2k^2 - (p\cdot k)^2]  \ G^{\rm as}((k + p)^2) 
\ G^{\rm as}(p^2) . }} 
Like the ghost equation \sdghostb, this equation is invariant under
renormalization because of the identity
$Z_g^2 Z_3 \tilde{Z}_3^2 = 1$, and we may again take all quantities to be
renormalized, with suppression of the index $r$.  We substitute the power-laws
\powerlaw\ into this equation. By the power counting argument that was used
for the ghost propagator, we again obtain the relation of the infrared
critical exponents
$\a_D + 2\a_G = - (4 - d)/2$.  This integral converges in the unltraviolet for
$d -2 < 4(1+\a_G)$, or $\a_G > (d - 2)/4$.  

	The gluon and ghost SD equations now read
\eqn\glugho{\eqalign{
(c_D c_G^2)^{-1} = I_D(\a_G) = I_G(\a_G),
}}
where
\eqn\sdgluonc{\eqalign{
I_D(\a_G) \equiv N & (2 \pi)^{-d} (d-1)^{-1}(k^2)^{-(2 + \a_D)}   \cr
& \times \int d^d p \  [p^2k^2 - (p\cdot k)^2]  \ [(k + p)^2]^{-(1 + \a_G)} 
\ [(p)^2]^{-(1 + \a_G)} 
}} 
\eqn\sdghostb{\eqalign{
I_G(\a_G) \equiv   
  N (2 \pi)^{-d} (p^2)^{-(1 + \a_G)}
\int d^d k \  & [p^2 k^2 - (p\cdot k)^2] 
   (k^2)^{-(2 + \a_D)}     \cr
& \times \{ \ (k^2)^{-(1 + \a_G)} - \ [(k + p)^2]^{-(1 + \a_G)} \},
}} 
and it is understood that $\a_D \equiv -2\a_G - (4-d)/2$.  The critical
exponent $\a_G$ is determined by the equality \glugho.  The integrals
$I_D(\a_G)$ and $I_G(\a_G)$ are evaluated in the Appendix, without angular
approximation, in arbitrary Euclidean dimension $d$.

\newsec{Determination of infrared critical exponents}

	To determine the critical exponent $\a_G$, we substitute the
formulas for
$I_D(\a_G)$ and $I_G(\a_G)$, given in the Appendix, into the equation
$I_D(\a_G) = I_G(\a_G)$ and obtain, for
$\a \equiv \a_G$,
\eqn\criteq{\eqalign{
f_d(\a) \equiv { {(d-1)\pi} \over {\sin(\pi \a)}} \ 
{ {\G(1 + 2\a)} \over {\G(- 2\a + d/2) \ \G(1 + \a + d/2)}}
\ { {\G(d - 2\a)} \over {\G(- \a + d/2) \ \G(1 + 2\a - d/2)}} = 1.
}}
We take the dimension $d$ of space-time in the interval 
$2 \leq d \leq 4$.  The integrals $I_D(\a)$ and $I_D(\a)$ are both
convergent in the ultraviolet only for $\a$ in the interval 
$ 0 < (d-2)/4 < \a < 1$, so the equation which determines $\a$ holds only in
this interval.   However whereas $I_D(\a)$ is manifestly positive
throughout this interval, the expression for
$I_G(\a)$ is negative for $\a > d/4$, because $1/\G(- 2\a + d/2)$, 
changes sign at $\a = d/4$,  so we look
for a solution only in the reduced interval
\eqn\redint{\eqalign{
0 \leq (d-2)/4 \leq \a \leq d/4 \leq 1.
}}
The identity, 
$\G(-2\a + d/2) \G(1 + 2\a - d/2) = \pi / \sin[\pi (-2\a + d/2)]$,
gives
\eqn\criteqa{\eqalign{
f_d(\a) \equiv { {(d-1) \sin[\pi(-2\a + d/2)]} \over {\sin(\pi \a)}} \ 
{ {\G(1 + 2\a)} \over { \G(1 + \a + d/2)}}
\ { {\G(d - 2\a)} \over {\G(- \a + d/2) }} = 1.
}}

	For the case of physical interest, $d = 4$, the allowed interval is 
$1/2 \leq \a \leq 1$, and the function $f_4(\a)$ contains the factor
$\sin[\pi(-2\a + 2)]/\sin(\pi \a) = \sin[\pi(-2\a)]/\sin(\pi \a)$
which is of the indeterminate form $0/0$ at $\a = 1$.  To control this (and a
similar indeterminacy for $d = 2$ at $\a = 0$), we first consider $d$ in the 
range $2 < d < 4$, and then take the limit $d\to 4$ (and $d \to 2)$.  For
$d$ in this range, one sees that 
the function $f_d(\a)$ is {\it positive} and finite, $f_d(\a) > 0$, for $\a$ in
the interior of  the allowed interval $(d-2)/4 < \a < d/4$, but vanishes at {\it
both} end-points $f_d[(d-2)/4] = f_d(d/4) = 0$ because of the factor 
$\sin[\pi(-2\a + d/2)]$.   It follows that the equation $f_d(\a) = 1$ has
an {\it even} number of solutions (if any)\foot{From numerical plots it
appears that for $2 < d < 4$ there are always 2 distinct real roots in the range
$(d-2)/4 < \a < d/4$, except possibly near $d \approx 2.662$ where there may be
a double root near $\a \approx 0.33095$.} for $2 < d < 4$.  We now set 
$d = 4$ and obtain,  
\eqn\criteqb{\eqalign{
f_4(\a) \equiv { {-3 \sin(2\pi \a)} \over {\sin(\pi \a)}} \ 
{ {\G(1 + 2\a) \ \G(4 - 2\a)} \over { \G(3 + \a) \ \G(2 - \a)} }
 = 1.
}}
We use
\eqn\identc{\eqalign{
\G(1 + 2\a) \ \G(4 - 2\a) & =  (3-2\a)(2-2\a)(1-2\a) \ 2\a \ 
\pi/\sin(2\pi \a)  \cr
 \G(3 + \a) \ \G(2 - \a) & = (2+\a)(1+\a)\a \ (1-\a) \pi / \sin(\pi \a),
}}
and obtain
\eqn\criteqb{\eqalign{
f_4(\a) = 12 \ { {(3 - 2\a)(2\a - 1)} \over {(2+\a)(1+\a)}} \ 
 = 1,
}}
where we have used 
${ { \sin(2\pi \a)} \over {\sin(\pi \a)}} 
\ { { \sin(\pi \a)} \over {\sin(2\pi \a)}} = 1$, which is valid only for 
$1/2 < \a < 1$.  This yields a quadratic equation with roots 
$\a = [93 \pm \sqrt(1201) ]/98 \approx [93 \pm 34.66 ]/98$.  Only {\it one}
root $\a \approx 0.5953$  lies in the interval $1/2 < \a < 1$.  On the other
hand we have just seen that for $2 < d < 4$, there are an {\it even}
number of roots.  The resolution is that for $d = 4 - \e$ there are two roots,
and the second root is given by $\a = 1 - O(\e)$, so in the limit 
$d \to 4$, there is a second root at $\a = 1$.

  We conclude that 
the infrared critical exponents $\a = \a_G$ and $\a_D = - 2\a_D - (4-d)/2$ are
given, in $d = 4$ dimensions, by two possible sets of values
\eqn\critexp{\eqalign{ 
\a_G & = 1; \ \ \ \ \ \ \ \ \ \ \ \ \ \ \ \ \ \ \ \ \ \ \ \ \ \ \ \ \ \ \ \ \
\ \ \ \ \ \ \ \ \ \ \ \ \ \a_D  = -2   \cr
\a_G & = [93 - \sqrt(1201) ]/98 \approx 0.5953
 \ \ \ \ \ \ \ \ \ \ \ \   
\a_D = - [93 - \sqrt(1201) ]/49 \approx - 1.1906.
}}
In the same way one finds for $d = 2$, 
\eqn\critexpa{\eqalign{ 
\a_G & = 0; \ \ \ \ \ \ \ \ \ \ \ \ \ \ \   
\a_D = -1    \cr
\a_G & = 1/5; \ \ \ \ \ \ \ \ \ \ \ \   
\a_D = - 7/5.
}}
For $d = 3$, one obtains
the equation, 
$f_3(\a) = { {32 \a (1-\a)[1 - \cot^2(\pi \a)]} \over {(3+2\a)(1+2\a)} } = 1$
with roots in the interval $1/4 \leq \a \leq 3/4$, given by 
\eqn\critexpb{\eqalign{ 
\a_G & = 1/2; \ \ \ \ \ \ \ \ \ \ \ \ \ \ \ \    
\a_D = -3/2    \cr
\a_G & \approx 0.3976; \ \ \ \ \ \ \ \ \ \ \ \   
\a_D \approx -1.2952.
}}
We expect that in each case one of the roots is spurious, and arises because
\criteqa\ does not express the full content of the theory.

	We note that in each case, one solution corresponds to 
$\a_G = (d-2)/2 = 0, 1/2,$ and 1, for $d = 2, 3,$ and 4, which gives
$G(k) \sim 1/(k^2)^{d/2}$.  This may be too infrared singular to be
acceptable.  But for $d = 2$, the other solution, with $\a_G = 1/5$,
is even more infrared singular, which suggests that for $d = 2$ the
first solution may be preferred namely, $\a_D = -1$ and $\a_G = 0$,
which may make the case $d = 2$ pathological in the Landau
gauge.  This case is exactly solvable in the axial gauge because the
non-linear term in the $d = 2$ Yang-Mills field is absent in this gauge,
and gives an area law at the classical level. There can of course be no
physical gluons in d = 1 + 1 dimensions even in the free theory which
may thus be considered confining.   Clearly the case $d = 2$ in the
Landau gauge requires a more detailed investigation that we do not
attempt here.

\newsec{Discussion and Conclusion}

	 We have seen that because the Faddeev-Popov weight $P_{FP}(A)$ contains nodal
surfaces, the SD equations corresponding to the Faddeev-Popov method are
ambiguous, and in practice one does not know how to select an exact
and globally correct solution.  Gribov's proposal, to cut-off the
Faddeev-Popov integral at the first nodal surface, produces a positive
probability distribution, but it is not exact because it overcounts some gauge
orbits, although it may give a useful approximation.  

	By contrast the method of stochastic quantization by-passes the
Gribov problem of selecting a single a representative in each gauge
orbit.  Instead the diffusion equation in A-space, eq.~\diffeqa,
contains an additional ``drift force" $a^{-1} D_\m \ \p \cdot A$ that
is a harmless generator of a gauge transformation.  The corresponding
DS equation that defines the non-perturbative Landau gauge was obtained
by solving the limit $a \to 0$ of this equation by the Born-Oppenheimer
method.   The limiting probability distribution $Q(A^{\rm tr})$ was
shown to vanish outside the Gribov horizon.  It is determined by a
diffusion equation that contains the novel term $K_2$, eq.~\avdrifta,
that corrects the Faddeev-Popov distribution cut-off at the Gribov
horizon for over-counting inside the Gribov horizon.

	[We may mention here an alternative approach.  The Landau gauge is the
singular limit $a \to 0$ of more regular gauges, and contains a
non-local effective drift force $K_{\rm gteff}$, eq. \avdrifta.  For
this reason it may be preferable to calculate with gauge parameter $a$
finite, so the drift force, 
$K_\m = - { { \d S_{\rm YM} } \over { \d A_\m } } + a^{-1}D_\m \p \cdot A$,
remains local, and there is no horizon outside of which the probability
distribution vanishes exactly.  In this case the SD equation \sdg\ for
the effective action $\Gamma$ gets replaced by
\eqn\sdga{\eqalign{
 \int d^4x \  {{ \d \Gamma(A) } \over {\d A_\m }}
\Big[ {{ \d \Gamma(A) } \over {\d A_\m}}
 +  K_\m \Big( A + {\cal D}(A){{ \d } \over {\d A}}\Big) \Big]  = 0.
 }}
The gluon propagator is given by $D^{-1} = R$,
where  
$R = - { {\d} \over {\d A} }K[A + {\cal D}(A) { {\d} \over {\d A} }]|_{A =
0}$, as in eq. \defq.  One would hope to solve the SD equations for the
full propagators in this approach, and not just their infrared
asymptotic limit.  An advantage of this approach is that the solution
for finite value of the gauge parameter $a$ could be directly compared
with the numerical lattice data of \nakamurac\ and \nakamurad\ that
is taken with stochastic gauge fixing and gauge parameter $a = 0.1$.  To
control ultraviolet divergences, it will be necessary to develop
Ward-type identities appropriate to this scheme.  They were not needed
in the present calculation because no ultraviolet divergences appeared
in the infrared limit.  Such identities in BRST-form are available in
the 5-dimensional scheme that is based on the time-dependent diffusion
equation
\zbgr, \bgz, \bulkqg, and  alternatively one may attempt to solve the SD
equations of the 5-dimensional scheme non-perturbatively.]

	In the second part of the article, where we calculated the infrared critical
exponents, we have however ignored the new term $K_2$ in order to compare with
other authors, and because it is not without interest to calculate the infrared
critical exponents non-perturbatively in Faddeev-Popov theory with a cut-off at
the Gribov horizon. 

	It is noteworthy that all our values for the critical exponents in $d =$ 2, 3
and 4 dimensions agree with exact results for a probability distribution that
is cut-off at the Gribov horizon namely, the vanishing \vanish\ of the
gluon propagator $D(k) \to 0$ as $k \to 0$,  and the
enhancement \horizcon, \horizpt\ and \czcoulf\ 
of the ghost propagator $[k^2 G(k)]^{-1} \to 0$, (except for the first
solution in $d = 2$ which is marginal, with $\a_G = 0$ and
$a_D = -1$).   The vanishing of $D(k)$ at $k = 0$ is counter-intuitive, and has
no other explanation than the proximity of the Gribov horizon in infrared
directions.  This suppresses the infrared components $A(k)$ of the gluon field,
and thus of the gluon propagator 
$D(k) = \langle |A(k)|^2 \rangle$.  Since our calculation involves a
truncation of the SD equations which is an uncontrolled approximation, 
the stability of our results should be tested by estimating
corrections.  	As for the future, an immediate challenge is to include
the effect of the new term $K_2$, eq.~\avdrifta, that was not evaluated
in the present calculation.  One must also introduce quarks. 

	We wish to compare our values of the infrared asymptotic dimensions with those
reported by \smekal, \atkinsona, and \atkinsonb.  But first we must
verify whether they also selected the solution of the SD equations of
Faddeev-Popov theory that vanishes outside the Gribov horizon.  Note that to
obtain a particular solution it is sufficient to require any one of its
properties, provided that this requirement selects a unique solution.  Indeed a
unique solution was obtained in~\smekal\ by requiring that both the gluon and
ghost propagators $D(k)$ and $G(k)$ be positive.  These properties by no means
follow from the Faddeev-Popov weight \fadpop\ that oscillates in sign, whereas
restriction to the Gribov region does imply the positivity of both $G(k)$ and
$D(k)$.  So in fact the restriction to the Gribov region is also implemented in
this way in~\smekal.  Likewise the assumptions made  in \atkinsona\ and
\atkinsonb\ to obtain a solution of the SD equations are equivalent to the
horizon condition, eq.~\horizony, that we imposed in sec.~6.

	It is reassuring that the values given in eq. \critexp\ for $d = 4$ 
agree qualitatively with the values reported in \smekal, namely 
$\a_G = [61 - \sqrt(1897)]/19 \approx 0.92$ and
$\a_D = -2 \a_G \approx - 1.84$, in the sense that the gluon propagator 
$D(k) \sim 1/(k^2)^{1 + a_D}$ vanishes at $k = 0$, and the ghost propagator
$G(k) \sim 1/(k^2)^{1 + a_G}$ is enhanced.  This may be an indication that
these qualitative features of the solution are not merely an artifact of the
approximations made.  For the two treatments of the SD equations are quite
different.  Indeed in \smekal, the gauge condition is treated off-shell, by
imposing the Slavnov-Taylor identities to determine longitudinal parts of
vertices, and by using the method of \brown\ to adjust the gluon propagator. 
On the other hand we have treated the gauge condition on-shell, so only
transverse quantities occur.  There is a similar qualitative agreement for $d
= 4$ with the values reported in \atkinsona, 
$a_G = (77 - \sqrt(2281))/38) \approx 0.769479$, and $a_G = -2 a_G$, where an
angular approximation was made.  The approximations made in
\atkinsonb\ appear to be similar to ours, although the method of solution is
quite different.  The value reported there for $d = 4$, $a_G = 1$ and $a_D =
-2$, agrees with our first solution.\foot{After completion of this article,
L. von Smekal has kindly informed me that the first value,
$\a_G = [93 - \sqrt(1201)]/98$ in $d = 4$
dimensions, was also obtained by C. Lerche \lerche.}

	We also wish to compare our results with numerical Monte Carlo
studies of propagators in Landau gauge.  Numerical gauge fixing 
to the Landau gauge is
achieved by minimizing, with respect to gauge transformations, the lattice
analog of  $F_A(g) =  \int d^d x \ |{^g}A|^2$, which indubitably produces
configurations that lie inside the Gribov horizon.  This gauge fixing,
like stochastic gauge fixing, has a Euclidean weight that is everywhere
positive, without over-counting.  However it is not in the class
of Faddeev-Popov gauges for which the determinent alternates in sign,
so a comparison with analytic calculations by the Faddeev-Popov method
does not have a completely clear interpretation.

	The infrared behavior of the lattice propagators is very sensitive to
finite-volume effects, and control of the volume dependence at fixed $\beta =
2N/g_0^2$ is required.  In particular $D(k)$ does not and should not
vanish at $k = 0$ at any finite lattice volume, but only when
extrapolated to infinite volume.   We have not attempted
here to fit the data of~\nakamurac\ and~\nakamurad\ without an estimate
of the finite-volume correction and the effect of the finite gauge
parameter, but this is a promising avenue for future comparison of
numerical and analytic results.  However we do note that it was
reported in \nakamurac, with stochastic gauge fixing at gauge parameter 
$a = 0.1$ (with Landau gauge at $a = 0$), that a fit          
to the Gribov formula, $D(k) = Z k^2[(k^2)^2 + M^4]^{-1}$,
(strong infrared suppression) can explain the gross feature of the
data.  Recent studies in Landau gauge at finite lattice volume indicate
a suppression of the gluon propagator in the infrared \bonnet, and are
not incompatible with an enhancement of the ghost propagator \suman.
The infrared behavior of the lattice gluon propagator $D(k)$ has been
studied in SU(2) gauge theory in Landau gauge in $d = 3$ Euclidean
dimensions~\acdland. It was found that $D(k)$ has a maximum at 
$k \approx 350$ MeV (normalized to the physical value of the string
tension) that is practically $\b$-independent, and that
$D(k)$ decreases as $k$ decreases below this value.  This decrease is
interpreted as resulting from the proximity of the Gribov horizon in
infrared directions.  A similar behavior is expected for the
3-dimensionally transverse part of the gluon propagator in Coulomb
gauge, in 4 Euclidean dimensions.  This has been observed, and an
extrapolation to infinite lattice volume at fixed~$\b$ was in fact
found, notably, to be consistent with the {\it vanishing} of 
$D(k)$ at $k = 0$~\cznumstgl\  and~\czfitgrib.  We emphasize that this
behavior is not seen at finite lattice volume but only in the
extrapolation to infinite lattice volume, at fixed $\beta$.  For
this reason it is important to extend the lattice calculations in
Landau, Coulomb and stochastic gauges to larger volumes, and to
extrapolate to infinite lattice volume before attempting a fit to
continuum formulas.  

	So what have we learned about propagators and the confinement problem
in QCD?  We may summarize results qualitatively by the statement that
in the infrared region in non-perturbative Landau gauge there is strong
suppression or vanishing of the would-be physical gluon propagator, and
strong enhancement of unphysical propagators.\foot{The ``unphysical
propagator" that is infrared-enhanced may be either the ghost
propagator in non-perturbative Landau-gauge Faddeev-Popov theory, or
the 44-component of the gluon propagator in Coulomb gauge.  The ghost
propagator coincides, approximately, with the remnant of the
longitudinal gluon propagator that survives the Landau-gauge limit in
stochastic quantization.}  This is
true both for the analytic solutions of the Schwinger-Dyson equations
obtained in \smekal, \smekrev, \atkinsonb, \lerche\ and here, and for
the numerical lattice data just discussed, with similar numerical data
in Coulomb gauge.  We expect that these qualitative features will stand
the test of time.  They provide a simple intuitive picture of
confinement in which the suppressed massless physical gluon disappears
from the physical spectrum while the enhanced unphysical components
provide a long-range color-confining force.  (This long-range force
should also confine quarks, but that has not been addressed here.)  As
discussed previously \critical\ both features may be understood as
the result of the restriction to the Gribov region, that results from
the identification of gauge-equivalent configurations.  The infrared
suppression of the transverse gluon propagator results from the
proximity of the Gribov horizon in infrared directions, while the
enhancement of the unphysical components is an entropy effect that
results from high population in the neighborhood of the Gribov horizon,
where the inverse Faddeev-Popov operator is enhanced.

\vskip .5cm
{\centerline{\bf Acknowledgments}}

It is a pleasure to thank Reinhard Alkofer, Alexander Rutenburg, Adrian
Seufert, Alan Sokal, and Lorenz von Smekal for valuable discussions. 
This research was partially supported by the National Science
Foundation under grant PHY-0099393.

\vskip .5cm

\appendix A{Evaluation of integrals}

	To evaluate the gluon self-energy $I_D(\a_G)$, eq. \sdgluonc, we write
\eqn\idx{\eqalign{
1/[(p-k)^2]^{1 + \a_G} 
= \G^{-1}(1 + \a_G)\int_0^\infty dx \ x^{\a_G} \exp[-x(p-k)^2],
}}
and similarly for $1/[(p)^2]^{1 + \a_G}$.  This gives
\eqn\sdgluond{\eqalign{
I_D(\a_G) \equiv N    [(d-1)(k^2)^{(2 + \a_D)}  \G^2(1 + \a_G)]^{-1}
\int_0^\infty dx \int_0^\infty dy \ (xy)^{\a_G} J
}}
where
\eqn\defj{\eqalign{
 J & \equiv (2 \pi)^{-d} \int d^d p
\  [p^2k^2 - (p\cdot k)^2]  \ \exp[-x(p-k)^2 - yp^2]  \cr
& = (d-1) \ k^2 
\ [2 \ (4\pi)^{d/2} \ (x+y)^{1 + d/2}]^{-1} \exp[ - (x+y)^{-1}xyk^2].
}} 
We introduce the identity $1 = \int d \g \ \d(x+y - \g)$ and change variable
according to $x = \g x'$ and $y = \g y'$.  This gives, after dropping 
primes,
\eqn\sdgluone{\eqalign{
I_D(\a_G) = N    (k^2)^{-(1 + \a_D)} 
[2 \ (4\pi)^{d/2} \ \G^2(1 + \a_G)]^{-1} K
}}
\eqn\defk{\eqalign{
K & \equiv\int_0^\infty dx \int_0^\infty dy \int_0^\infty d\g \ \d(x+y-1)
\ (xy)^{\a_G} \ \g^{2\a_G - d/2} \exp[- \g xyk^2]   \cr
& = (k^2)^{-2\a_G -1 + d/2} 
\ { { \G(2\a_G + 1 - d/2) \ \G^2(-\a_G + d/2)} \over {\G(d - 2\a_G)} }.
}}
This gives
\eqn\sdgluonf{\eqalign{
I_D(\a_G) = { {N} \over {2 \ (4\pi)^{d/2} } }  
\ { {\G(2\a_G + 1 - d/2) \ \G^2(-\a_G + d/2)}
 \over {\G^2(1 + \a_G)\G(d - 2\a_G)} },
}}
where we used $\a_D = - 2 \a_G - (4 - d)/2$.  

	To evaluate the ghost self-energy, $I_G(\a_G)$, eq. \sdghostb, we use the
identities
\eqn\idgh{\eqalign{
{ {1}  \over {[(k)^2]^{2 + \a_D} } } 
= { {1}  \over { \G(2 + \a_D) } } 
\int_0^\infty dx \ x^{\a_D + 1} \exp(-xk^2),
}}
\eqn\idghb{\eqalign{
{ {1}  \over {[(k)^2]^{1 + \a_G} } } 
-  { {1}  \over {[(k - p)^2]^{1 + \a_G} } }  
= & \ { {1} \over {\G(1 + \a_G)} }  \int_0^\infty dy \ y^{\a_G + 1} 
 \ [(k-p)^2 - k^2]   \cr
& \times   \int_0^1 dz 
 \ \exp\{-y[z(k-p)^2 + (1-z)k^2]\} , 
}}
which allow us to cancel the leading power of $k$ explicitly.  This
gives
\eqn\sdghostc{\eqalign{
I_G(\a_G) =   
  &   \ { {N} \over { \ (p^2)^{1 + \a_G} \ \G(2 + \a_D) \ \G(1 + \a_G)} }  
    \cr
&  \ \ \ \ \ \ \ \ \ \ \ \ \times \int_0^\infty dx  \int_0^\infty dy 
    \int_0^1 dz \ x^{\a_D + 1} \ y^{\a_G + 1} \ L   
}}
where
\eqn\idl{\eqalign{
L \equiv   
   \ (2 \pi)^{-d} 
\int d^d k & \  [p^2 k^2 - (p\cdot k)^2] (p^2 - 2p \cdot k)    \cr
& \times    \ \exp\{- [x + y(1-z)]k^2 - yz(k-p)^2 \}   . 
}}
\eqn\idla{\eqalign{
L =  { {(p^2)^2 \ (d-1)} \over {2 \ (4 \pi)^{d/2} } } \ 
 { {x + y - 2 yz} \over {(x+y)^{2 + d/2} } }    
    \ \exp\Big(- { {yz[x + y(1-z)]p^2} \over {x+y} } \Big) . 
}}
This gives
\eqn\sdghostd{\eqalign{
I_G(\a_G) =   
     \ { {N \ (d-1) \ (p^2)^{1 - \a_G}} 
\over { 2 \ (4 \pi)^{d/2} \ \G(2 + \a_D) \ \G(1 + \a_G)} } \ J
}}
\eqn\identj{\eqalign{    
J \equiv \int_0^\infty dx  \int_0^\infty dy 
    \int_0^1 dz \ x^{\a_D + 1} \ y^{\a_G + 1} 
& \ { {x + y - 2 yz} \over {(x+y)^{2 + d/2} } }     \cr
& \times \ \exp\Big(- { {yz[x + y(1-z)]p^2} \over {x+y} } \Big) .    
}}
We again introduce the identity $1 = \int d \g \ \d(x+y - \g)$ and change
variables according to $x = \g x'$ and $y = \g y'$.  This gives, after dropping 
primes,
\eqn\identja{\eqalign{    
J = \int_0^\infty d\g \int_0^\infty dx  \int_0^\infty dy 
    \int_0^1 dz & \ \d(x+y-1) \ x^{\a_D + 1} \ y^{\a_G + 1} \ (1 - 2yz)
     \cr
& \times \ \g^{-\a_G} \ \exp\Big(- yz[x + y(1-z)] \g p^2 \Big) ,    
}}
\eqn\identjb{\eqalign{    
J = (p^2)^{\a_G - 1} \G(1-\a_G)
  \int_0^1 dy 
    \int_0^1 dz & \ y^{\a_G + 1} \ (1-y)^{\a_D + 1}  \ (1 - 2yz)    \cr
& \times [yz(1 - yz)]^{\a_G-1} ,    
}}
where we again used $\a_D = - 2 \a_G - (4 - d)/2$.  We change variable of
integration to $u = yz(1-yz)$, with $du =y(1-2yz)dz$, and obtain
\eqn\identjb{\eqalign{    
J & =  (p^2)^{\a_G - 1} \ \a_G^{-1} \ \G(1-\a_G)
  \int_0^1 dy \ y^{2\a_G} \ (1-y)^{ - \a_G - 1 + d/2} \cr
	&  = { {(p^2)^{\a_G - 1} \  \G(1-\a_G) } \over {\a_G} } 
	\    { {\G(1 + 2\a_G) \ \G( - \a_G + d/2)}  \over { \G(\a_G + 1 + d/2)} }, 
}}
where we have again used $\a_D = - 2 \a_G - (4 - d)/2$. This gives
\eqn\sdghoste{\eqalign{
I_G(\a_G) =   
     \ { {N \ (d-1) } 
\over { 2 \ (4 \pi)^{d/2} } }
&  { {\pi} \over {\sin(\pi \a_G)} }    \cr
 & \times \ { {\G(2\a_G + 1) \ \G(- \a_G + d/2)} 
  \over {\G^2(\a_G + 1) \ \G(- 2\a_G + d/2) \ \G(\a_G + 1 + d/2)} }  ,
}}
where we have used $\G(\a_G) \ \G(1 - \a_G) = \pi / \sin(\pi \a_G)$.
This integral is positive for $\a_G < d/4$.

\footatend\vfill\supereject\immediate\closeout\rfile\writestoppt
\baselineskip=14pt\centerline{{\bf References}}\bigskip{\frenchspacing%
\parindent=20pt\escapechar=` \input refs.tmp\vfill\eject}\nonfrenchspacing

%%%%%%%%%%%%%%%%%%%%%%%%%%%%%%%%%%%%%%%%%%%%%%%%%%%%%%%%%%%%%%%%%%%
  %  \listrefs

\bye